\documentclass[lettersize,journal]{IEEEtran}
\usepackage{amsmath,amsfonts}
\usepackage{algorithmic}
\usepackage{algorithm}
\usepackage{array}
\usepackage{textcomp}
\usepackage{stfloats}
\usepackage{url}
\usepackage{verbatim}
\usepackage{graphicx}
\usepackage{cite}
\usepackage{multirow}
\usepackage{xcolor}
\usepackage[font=footnotesize,labelformat=simple]{subcaption}

\usepackage{cleveref}
\usepackage{soul}
\usepackage{amsthm}
\newtheorem{remark}{Remark}

\newcommand{\lightgray}[1]{{\color{lightgray} {#1}}}
\begin{document}

\title{Unlocking Realism and Interpretability in Wireless Channel Synthesis: A Physics-Guided Generative Approach}
\author{Satyavrat Wagle, Akshay Malhotra, Shahab Hamidi-Rad, Aditya Sant, David J. Love, Christopher G. Brinton}
\author{
\IEEEauthorblockN{Satyavrat Wagle$^1$, Akshay Malhotra$^2$, Shahab Hamidi-Rad$^2$, Aditya Sant$^2$, David J. Love$^1$, and Christopher G. Brinton$^1$}

\IEEEauthorblockA{$^1$Elmore Family School of Electrical and Computer Engineering, Purdue University, $^2$Interdigital Communications}
}

\maketitle

\begin{abstract}
    In recent years, machine learning (ML) methods have become increasingly popular for wireless communication systems.
    These require large amounts of data reflecting the behavior of realistic channels with high fidelity. However, sampling over-the-air (OTA) channel data is an extremely resource-intensive process which cannot accurately represent the variety of real world channels. This results in the need for realistic training data for ML systems. 
    To this end, generative models have been proposed to synthesize channel data. However, \textit{(i)} the outputs produced by such methods may not correspond to physically viable channels, \textit{(ii)} the outputs may not provide insights into the associated environment, and \textit{(iii)} training the generative model may need labeled data, requiring resource intensive data annotation.
    Through this work, we address these issues by integrating a parametric, physics-based geometric channel (PPGC) modeling framework derived from planar wave propagation equations, with generative methods to produce realistic channel matrices with interpretable representations in the parameter domain.
    To overcome the limitations of the resulting non-convex optimization landscape, we propose a linearized reformulation of the PPGC model to ensure smooth gradient flow during training, while also providing insights into the underlying physical environment. We incorporate a tensor decomposition framework into the linearized reformulation to allow for flexibility in the number of wireless channel parameters. We also show the compatibility of this reformulation with parameter extraction tasks. 
    We evaluate our model against prior baselines by comparing generated, scenario-specific samples to true channels in terms of their similarity and through their utility in downstream compression tasks. \let\thefootnote\relax\footnote{This work was completed during Satyavrat Wagle's internship at InterDigital Communications.   Prof. Christopher G. Brinton and David J. Love's contributions to this article were supported by the Office of Naval Research (ONR) grant N00014-21-1-2472, and the National Science Foundation (NSF) grants CNS-2212565 and ITE-2326898. Prof. Christopher G. Brinton was also supported in part by the Defense Advanced Research Projects Agency (DARPA) grant D22AP00168 and by the Air Force Office of Scientific Research (AFOSR) under grant FA9550-24-1-0083.}
\end{abstract}

\section{Introduction}\label{sec:intro}

\IEEEPARstart{T}he use of machine learning (ML) for applications in signal processing and wireless communication has seen extensive interest in the past few years. At the physical layer (PHY) of wireless communication systems, ML research has predominantly focused on two main objectives: estimating and mitigating distortions in electromagnetic signals during over-the-air (OTA) transmission (such as channel compression, estimation, equalization, and beamforming),\cite{csinet,ch_estimation,beamforming,key_areas_in_6g} and addressing noise and non-linearities at the antennas \cite{estimation_survey,sant2024insights}.

However, despite being a promising solution for PHY layer communication tasks, ML methods in these domains are frequently outperformed by traditional model-based methods \cite{model_based_1,model_based_2,model_based_3}, because of the resource-intensive tasks of manually collecting, cleaning and labeling wireless over-the-air (OTA) data from the real world
to effectively train ML models for practical PHY layer deployments. Past data measurement and labeling campaigns have taken multiple months to capture a small number of fully characterized data points for a single scenario \cite{OTAdata_capture1,OTAdata_capture2,OTAdata_capture3}, while ML pipelines typically require substantial amounts of training data. Additionally, the full diversity of the wireless channel cannot be captured by OTA data collection due to hardware limitations and dynamism in the environment of interest \cite{ota_diversity_1,ota_diversity_2}. These drawbacks have a detrimental impact on downstream wireless ML systems which rely on diverse and realistic channel data for training \cite{ota_perf_1,ota_perf_2}. Modern ray-tracing packages and software \cite{sionna} replicate diverse wireless environments. However, these models require extensive computational resources in creating channel data, limiting their use for online training on low-power devices like IoT devices. Further, these models cannot be easily tailored to generate channel models from a particular distribution in the channel parameter space. In contrast, this work proposes a scalable yet interpretable PPGC-based generative model. As the model learns channel generation through parameter generation, the generated channels naturally correspond to a particular distribution in the parameter space.

Next, we first provide an overview of the different generative channel modeling approaches, followed by a brief description of the contributions of this work.

\subsection{Related Work}
The use of generative ML models has been proposed to artificially synthesize wireless data \cite{GAN1} to mitigate the problems described in Sec. \ref{sec:intro}. Generative models aim to learn the underlying distribution of data by learning either its parametrized statistical model \cite{parametric_genai]}, or a mapping from a known, simple distribution to the distribution of interest \cite{implicit_genai}. 
A generative  adversarial network (GAN) based wireless channel modeling framework was first introduced in \cite{GAN1}. In \cite{channelgan}, the authors utilize a Wasserstein-GAN with Gradient Penalty (WGAN-GP) to synthesize novel channel matrices, which were evaluated by cross-validating between real and synthesized data. In \cite{mimogan}, the authors trained a GAN on multiple-input multiple-output (MIMO) data, explicitly designed to learn the spatial correlation across the channel data. In \cite{diffusion_models_for_channels}, a diffusion based generative model has been adopted to circumvent the issue of mode collapse in GANs. 
Works such as \cite{score_based} utilize a score-based generative model to generate channel matrices and for channel estimation in noisy environments. In \cite{cond_gan}, the authors train a conditional GAN to generate environment-specific channel matrices by conditioning the generated samples on the RF signatures of the environment. In \cite{irs_gan}, the authors leverage prior knowledge of the reflected channel to generate novel samples of channel matrices for intelligent reflecting surface-equipped systems. In \cite{bock}, the authors utilize received pilot signal data to predict the distribution of the wireless channel.

A similar research direction involves using labeled datasets to predict parameters associated with the wireless channel. 
In \cite{gen_models_mmwave_uav}, the authors use the locations of UAVs to predict the link state and the parameters sequentially, using a conditional variational autoencoder (VAE) to capture relationships within the data. 
Similarly, in \cite{multi_freq_model} the authors develop a GAN-based model to generate new instances of parameters given the location of UAVs. 
In \cite{conv_gan}, a convolutional GAN is used to produce time-varying, frequency-selective channels. 
In \cite{r21,r22}, the authors capture channel characteristics through scatterer-based modeling of the propagation environment as a low-complexity alternative to ray tracing based channel synthesis.

Now, unlike common modalities of data that we typically encounter, such as image, text, audio, etc. which are directly human-interpretable, the wireless channel data is a tensor of complex numbers and is \emph{not human-interpretable or easily visualized}. This gives rise to two challenges unique to the usage of generative models for wireless channel generation.
Firstly, while the generative models are trained to generate data points statistically similar to the training set, owing to the stochasticity of the model, the synthesized outputs may not correspond to valid channels.  Here, the validity of channel data implies that the wireless channel can be represented as a multipath geometric model representing the multiple paths the transmitted signal takes, before arriving at the receiver \cite{chanmodel_1,chanmodel_2}. Therefore, there are no guarantees about the quality of generated data, which can potentially have adverse effects on downstream ML tasks \cite{genai_for_wireless_2,genai_for_wireless_3}.
Secondly, it is hard to gain any insights about the physical parameters associated with the signal propagation or any information about the environment or scenario being considered (e.g. angles associated with paths, gains of paths, line-of-sight transmission or non-line-of-sight, etc.) from generated data samples. This lack of interpretability hinders the ability to evaluate the efficacy of the generative model to synthesize context-specific data \cite{genai_for_wireless_3}. 


\subsection{Proposed Solution and Contributions}
Our proposed method to generate the wireless channel overcomes the aforementioned challenges by incorporating a verified PPGC model into the generative pipeline. As the PPGC model parametrizes the channel generation, our generative process learns the joint distribution of the underlying parameters responsible for channel generation. 
This mitigates both the aforementioned issues of prior approaches for generative channel modeling. Firstly, incorporating a verified PPGC model in the pipeline guarantees that the outputs of our framework are valid channels, and secondly, the model generates the parameters associated with the channels, ensuring we can extract insights related to the physics of the environment, making the generated channels more interpretable. 


In contrast to the works in \cite{channelgan,mimogan,diffusion_models_for_channels,score_based,cond_gan,irs_gan}, which use a generative model to directly produce channels without any guarantees on the validity of the generated matrices, our method uses the generative model to produce a distribution over the parameter domain which is utilized by a verified channel model to produce channel matrices guaranteed to be geometrically valid. Compared to \cite{bock}, which uses responses to pilot signals for single-input multiple-output (SIMO) data and considers the azimuth plane, we consider a MIMO system in both the azimuth and elevation planes, and do not require the knowledge of pilot signals. Our approach also provides mechanisms to explicitly extract parameter values, as well as to adapt to additional parameters used to model the channel. 
Compared to \cite{gen_models_mmwave_uav,multi_freq_model,conv_gan, r21, r22}, which require datasets labeled with metadata relating to the environment, locations of device and the entire set of channel parameters, our method does not require labeled data or prior knowledge and can learn channel parameters directly from channel matrices acquired through standard 3GPP CSI compliant pipeline post channel estimation at the user device, or at the network side via CSI reporting \cite{3gpp_data_collection}. 

A summary of the contributions of this work is as follows: 
\begin{itemize}
    \item We propose a generative ML framework which leverages a parametric, physics-based channel (PPGC) model to produce realistic channel data that belongs to the distribution of interest and illustrate the convergence related challenges arising from the non-convexity of the PPGC model when training the generative pipeline (Sec. \ref{sec:pred_params}). 
    \item We develop a linearized relaxation of the PPGC model to mitigate the effect of this non-convexity using a discretized dictionary of array response matrices across the range of parameter values (Sec. \ref{sec:pred_matrix}).
    \item In order to better scale the generative channel parameter estimation with the number of channel parameters, we incorporate a tensor decomposition method, leveraging the sparsity in the generator outputs (Sec. \ref{sec:tensor_decomp}).
    \item We design a parameter estimation module to calculate the parameters associated with a generated sample of data by leveraging the sparsity characteristics of the ML model outputs using peak detection algorithms (Sec. \ref{sec:parameter_estimation}).
    \item We show that our method can accurately generate channel data as well as the parameter distributions associated with a given set of real data. This is evaluated against prior art baselines on practical downstream tasks. We also show that our model can accurately estimate the underlying parameters for input channel matrices (Sec. \ref{sec:experiments}).
\end{itemize}

An abridged version of this work appeared in \cite{wagle2025physicsbased} and \cite{wagle_iclr}.
In this extension, we make the following significant additions to \cite{wagle2025physicsbased,wagle_iclr}: (1) we extend our method to the azimuth and elevation planes, and (2) incorporate tensor decomposition methods in the generative pipeline by leveraging the sparsity of the gain tensor, mitigating the issues of scaling to larger resolutions. (3) We develop a PPGC-aided parameter extraction framework to calculate path parameters from a given channel; and (4) we significantly expand upon our experimental results by evaluating our method against a larger set of baselines, demonstrating its effectiveness in parameter extraction tasks and flexibility to incorporate additional channel parameters.

\section{System Model and Approach}\label{sec:model}

\begin{figure}
    \centering
    \includegraphics[trim={2.7cm 0 3.3cm 0},clip,width=0.98\columnwidth]{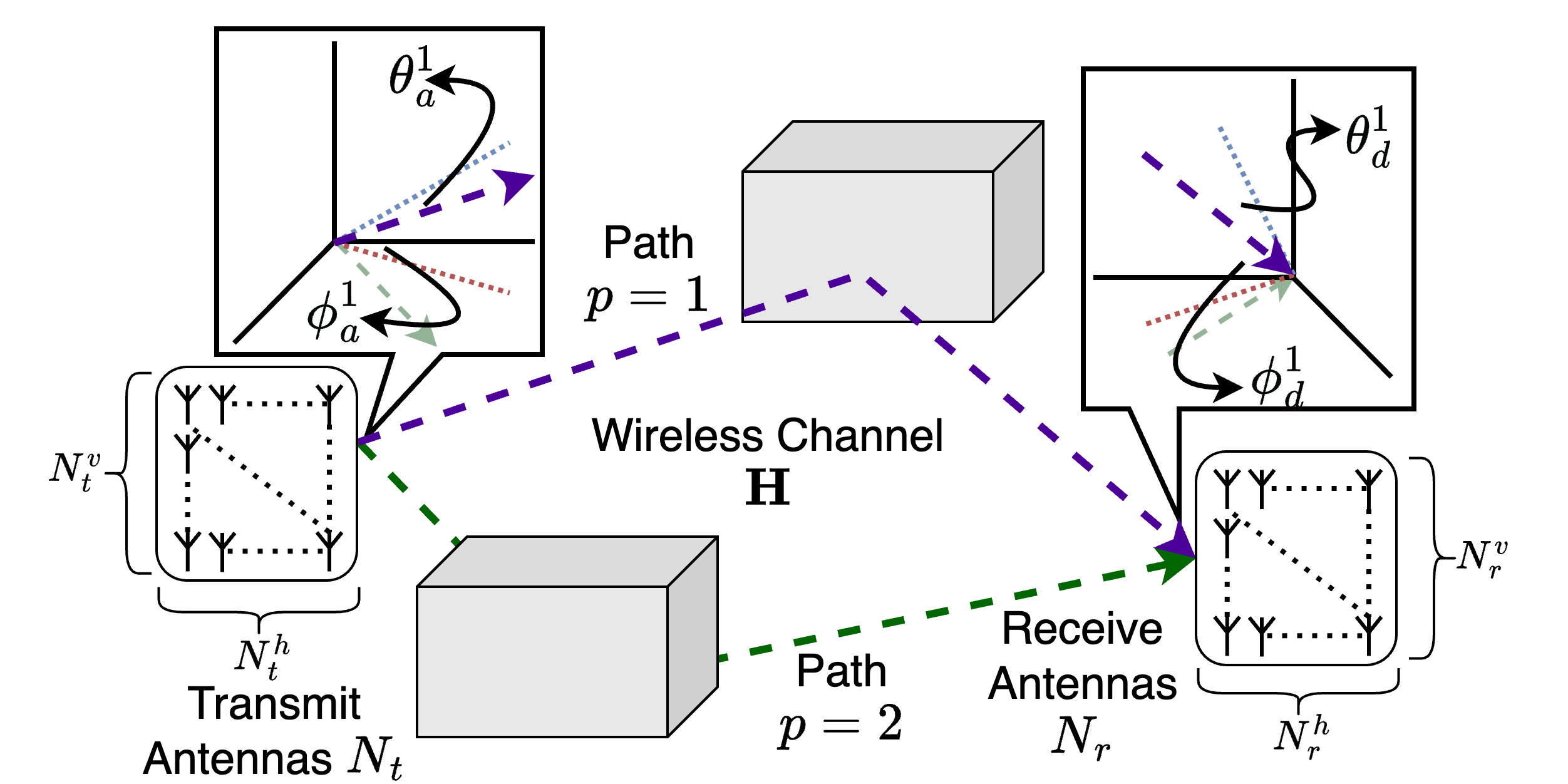}
    \caption{The physics-based geometric channel (PPGC) model is motivated by ray-tracing methods for channel tracking, and uses the arrival and departure angles in the elevation and azimuth plane for $P$ paths and the associated gains as parameters to model the wireless channel $\textbf{H}$.}
    \label{fig:comm_model}
\end{figure}

\begin{figure*}
    \centering
    \includegraphics[trim={0 65cm 88cm 0},clip,width=0.45\textwidth]{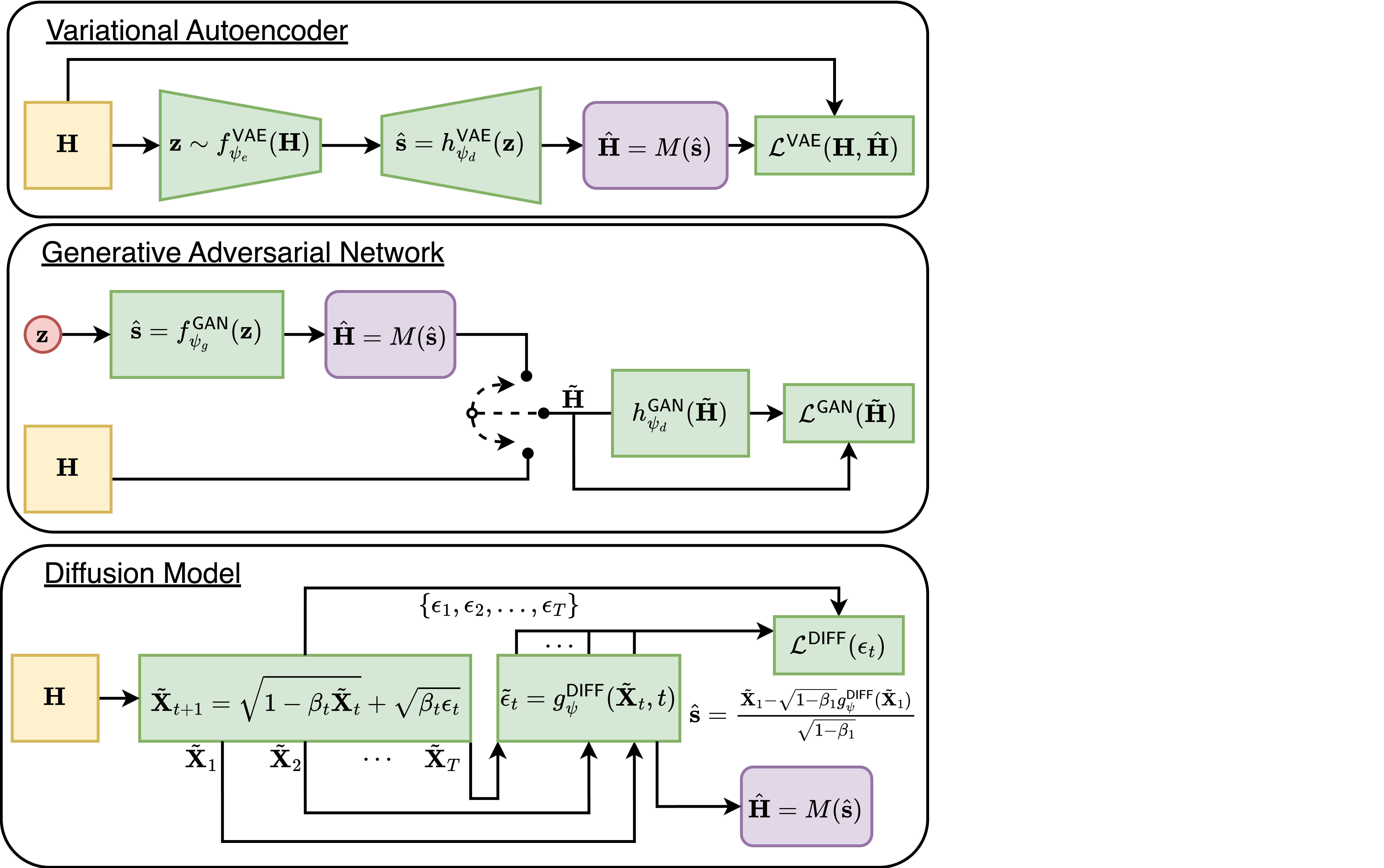}
    \includegraphics[width=0.5\textwidth]{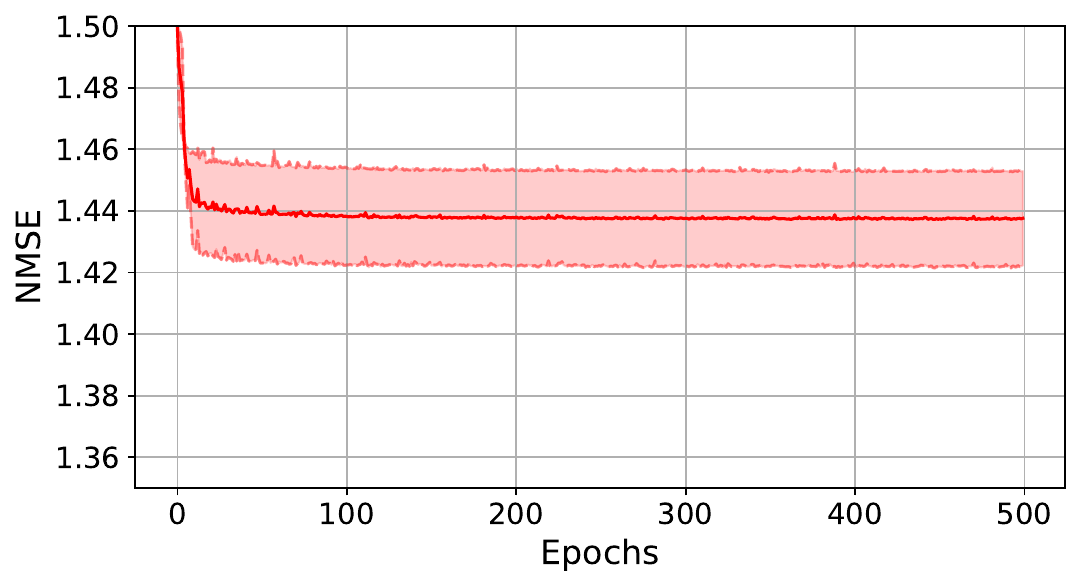}
    \caption{When incorporating the PPGC model $\mathcal{M}$ into the generative pipeline in a straightforward manner, the generator directly predicts the parameters $\hat{\textbf{s}}$, which are then used by the model $\mathcal{M}$ to predict the channel. Such a pipeline is compatible with architectures that do not require knowledge of the target generator outputs, with the generator-specific processes to produce the parameter vector $\hat{\textbf{s}}$ (Left). In such cases, the generator cannot converge to suitable optima due to the non-convexity of the PPGC model, as observed in the stagnation of training performance of the pipeline over multiple training runs (Right).}
    \label{fig:model_train}
\end{figure*}

In this section, we describe our proposed system. We begin by elaborating on the PPGC model that is used in the generative process, followed by the design of the generative model. We then discuss the design choices made for the generative model. Finally, we describe the training and inference pipeline.

\subsection{Channel Model}\label{sec:channel_model}

For MIMO systems, the wireless channel is expressed as a complex matrix $\textbf{H} \in \mathbb{C}^{N_t \times N_r}$, where $N_t, N_r$ are the number of transmit and receive antennas respectively, such that $N_t = N_t^h\cdot N_t^v$ and $N_t = N_r^h\cdot N_r^v$, where $N^h_{t/r},N^v_{t/r}$ are the number of antenna elements in each row and column respectively, as shown in Fig. \ref{fig:comm_model}. We assume a narrowband channel \cite{nbchannel_rapaport}, such that the signal received over the channel at discrete-time $k$ is only dependent on the waveform transmitted at time $k$, as a majority of modern communication systems leverage wideband OFDM communication, which can be interpreted as multiple narrowband channels due to the orthogonality of the channels corresponding to each subcarrier \cite{3gpp_ts38211,nbchannel_rapaport}. An established method of modeling the wireless channel $\textbf{H}$ is by using ray-tracing algorithms \cite{ray_tracing_1,ray_tracing_2}, wherein the path of the $p$-th individual ray is geometrically modeled to calculate the channel parameters such as angles of arrival and departure for the elevation plane ($\theta_a^p,\theta_d^p$), azimuth plane ($\phi_a^p,\phi_d^p$), and the propagation gain $g^p$ using the Saleh-Valenzuela PPGC model \cite{sv_model_1,sv_model_2} as shown in Fig. \ref{fig:comm_model}. Through 
this planar wave model, the contribution of the $p^{\mathrm{th}}$ path to the wireless channel is characterized by the response matrices $\textbf{A}_r(\theta_a^p,\phi_a^p) \in \mathbb{C}^{N_r^h \times N_r^v}$ and $\textbf{A}_t(\theta_d^p,\phi_d^p) \in \mathbb{C}^{N_t^h \times N_t^v}$, scaled by the complex gain $g^{p}$ \cite{rt_multipath}. We consider Uniform Planar Array (UPA) \cite{chanmodel_1} antennas, whose array response matrices for the transmit and receive antennas can be defined as
\begin{align}
\label{eq:array_responses}
    \textbf{A}_t(\theta_d^p,\phi_d^p) = \left[\frac{e^{ju\textbf{v}_{m,n}(\theta_d^p,\phi_d^p)}}{\sqrt{N_t}}\right]_{m=0,n=0}^{m=N^h_t-1,n=N^v_t-1},\\
    \textbf{A}_r(\theta_a^p,\phi_a^p) = \left[\frac{e^{ju\textbf{v}_{m,n}(\theta_a^p,\phi_a^p))}}{\sqrt{N_r}}\right]_{m=0,n=0}^{m=N^h_r-1,n=N^v_r-1},
\end{align}
respectively, where $\textbf{v}_{m,n}(\theta,\phi) = m \cdot \cos(\phi)\sin(\theta)+ n \cdot \cos(\theta)$ and $u = \frac{2\pi}{\lambda}d$; $\lambda$ is the wavelength of the carrier and $d$ is the distance between antenna elements. Thus, the PPGC model given by $\mathcal{M}: \mathbb{R}^{5P} \rightarrow \mathbb{R}^{2 \times N_t \times N_r}$, to map a set of parameters $\textbf{s} = [g_p,\theta_a^p,\theta_d^p, \phi_a^p, \phi_d^p]_{p = 1}^P \in \mathbb{R}^{5P}$ to a matrix $\textbf{H} \in \mathbb{C}^{N_t \times N_r}$ as
\begin{equation}
\label{eq:channel_sum}
    \textbf{H} = \mathcal{M}(\textbf{s}) = \sum_{p=1}^{P} g_p \textbf{A}_r(\theta^p_a, \phi^p_a) \textbf{A}_t(\theta^p_d, \phi^p_d)^H,
\end{equation}
where $P$ is the number of paths that a transmitted signal takes before being received at the receiver antennas, which is typically small in over-the-air transmission, such as in mmWave communication, where multipath sparsity is caused by severe path loss in the propagation environment, as discussed in \cite{rappaport2013millimeter,rangan2014millimeter,xing2021indoor}. All angles take values between $[-\pi,\pi]$ radians.




Now, the calculated channel $\textbf{H}$ differs significantly based on the physics of the environment \cite{ota_diversity_1,ota_diversity_2}, which relates to the parameters $\textbf{s}$ associated with $\textbf{H}$. Thus, a key characteristic of the PPGC model $\mathcal{M}$ is that, a distribution $q(\cdot)$ over the parameters $\textbf{s}$ induces a distribution $q_M(\cdot)$ over the channel matrix $\textbf{H}$, where the distribution $q(\cdot)$ is specific to the environment in which the model is deployed. In order to incorporate the PPGC model into the generative pipeline, the generative model must learn the prior distribution over the parameters $\textbf{s}$ instead of directly learning the prior distribution over the channel matrix $\textbf{H}$. Thus, the objective of our system is as follows; given a set of channel matrices $\mathcal{D} = [\textbf{H}_i]_{i=1}^N$, where $\textbf{H}_i \sim q'(\cdot)$ and $q'(\cdot)$ is unknown, we train a generative model to output a probability distribution $q(\cdot)$ over the parameters $\textbf{s}$ such that the induced distribution $q_M(\cdot)$ over the channel $\textbf{H}$ is close to $q'(\cdot)$.


\subsection{Generative Model to Predict Channel Statistics} \label{sec:pred_params}

We consider a generative model $g_{\psi}$ parametrized by $\psi$,  which operates in two modes, namely training mode, where the process $g^{\textsf{Train}}_{\psi}:\mathbb{R}^{2 \times N_t \times N_r}\rightarrow\mathbb{R}^{5P}$ takes a channel matrix $\textbf{H}$ as input and produces a parameter vector $\hat{\textbf{s}}$ as output as $\hat{\textbf{s}} = g_{\psi}^{\textsf{Train}}(\textbf{H})$, and generation mode, where the process $g^{\textsf{Gen}}_{\psi}:\mathbb{R}^{Z}\rightarrow\mathbb{R}^{5P}$ takes a sample $\textbf{z} \sim \pi_Z(\cdot) \in \mathbb{R}^{Z}$ from a known distribution $\pi_Z(\cdot)$ as input and produces a generated parameter vector $\hat{\textbf{s}}$ as output as $\hat{\textbf{s}} = g_{\psi}^{\textsf{Gen}}(\textbf{z})$. For both modes, the parameter vector $\hat{\textbf{s}}$ is then passed to the PPGC model $\mathcal{M}$ to produce a valid channel matrix as $\hat{\textbf{H}} = \mathcal{M}(\hat{\textbf{s}})$. 


\subsubsection{Generative model training}

We assume that the loss function of the generative model is a function of the input channel matrix and the output of the complete generative pipeline $\mathcal{M}(g_{\psi}^{\textsf{Train}}(\textbf{H}))$, and does not require knowledge of the target outputs of the generative model $g_{\psi}^{\textsf{Train}}(\textbf{H})$. The model architectures that satisfy this condition are variational autoencoders (VAE) \cite{vae} and generative adversarial networks (GAN) \cite{GAN1}. The particular processes by which the generative models produce the parameter vectors $\hat{\textbf{s}}$ and by which they are updated vary based on the nature of the generative model. We will empirically illustrate the compatibility of the generative pipeline with these models in Sec. \ref{sec:experiments}. In contrast, methods such as normalizing flow models \cite{norm_flows} assume that the transformation pipeline is invertible, which is incompatible with the modifications made by our method that produces rank deficient outputs, as we will explain in Sec. \ref{sec:tensor_decomp}, and diffusion models \cite{diffusion_models} require knowledge of the target generator outputs, as they operate by successively denoising the latent distribution.

For VAEs, which consist of encoder and decoder models given by $f^{\textsf{VAE}}_{\psi_e}$ and $h^{\textsf{VAE}}_{\psi_d}$ respectively (i.e. $g_{\psi} = \{f^{\textsf{VAE}}_{\psi_e}, h^{\textsf{VAE}}_{\psi_d}\}$), the encoder first samples a latent vector from the posterior distribution as $\textbf{z} = f_{\psi_e}(\textbf{H})$, which is then used to produce a parameter vector as  $\hat{\textbf{s}} = h_{\psi_e}(\textbf{z})$. The VAE is trained to map the unit Gaussian distribution to the latent distribution over the parameters, enabling the generation of novel parameter vectors $\hat{\textbf{s}}$. Valid channels are obtained as $\hat{\textbf{H}} = \mathcal{M}(\hat{\textbf{s}})$. The VAE loss is a generalization of the evidence based lower bound (ELBO)\cite{vae}, given by $\mathcal{L}^{\textsf{VAE}} = ||\textbf{H}-\hat{\textbf{H}}||_2^2 + \alpha_{D} \cdot \textsf{KL}(\textbf{z},\mathcal{N}(0,\textbf{I}))$. Here, the first term ensures that the outputs are similar to the inputs. The second term encourages the distribution of the latent vectors to be similar to $\mathcal{N}(0,\textbf{I})$. Thus, for VAEs, $g_{\psi}^{\textsf{Train}}(\textbf{H}) = f^{\textsf{VAE}}_{\psi_e}(h^{\textsf{VAE}}_{\psi_d}(\textbf{H}))$, and $g_{\psi}^{\textsf{Gen}}(\textbf{z}) = h^{\textsf{VAE}}_{\psi_d}(\textbf{z})$, where $\textbf{z}$ is sampled from a Gaussian distribution defined by $\mathcal{N}(0,\textbf{I})$.


GANs consist of a generator $f^{\textsf{GAN}}_{\psi_g}$ which learns to produce the parameter vector $\hat{\textbf{s}}$ from random noise, and a discriminator $h^{\textsf{GAN}}_{\psi_d}$ which learns to distinguish between real and fake channel data (i.e. $g_{\psi} = \{f^{\textsf{GAN}}_{\psi_g}, h^{\textsf{GAN}}_{\psi_d}\}$). The GAN is trained until $h^{\textsf{GAN}}_{\psi_d}$ cannot distinguish between ground truth and fakes produced by $f^{\textsf{GAN}}_{\psi_g}$, using min-max optimization, which is characterized by $\mathcal{L}^{\textsf{GAN}}=-\mathbb{E}_{\textbf{H}\sim\mathcal{D}_{\textbf{H}}}[\log h^{\textsf{GAN}}_{\psi_d}(\textbf{H})] -\mathbb{E}_{\mathbf{z}\sim p_Z(\mathbf{z})}[\log h^{\textsf{GAN}}_{\psi_d}(\mathcal{M}(f^{\textsf{GAN}}_{\psi_g}(\textbf{z})))]$. Here, $\mathcal{D}_{\textbf{H}}$ and $p_Z(\cdot)$ are the training dataset of valid channels and a known random distribution, respectively. Thus, for GANs, $g_{\psi}^{\textsf{Train}}(\textbf{H}) = h^{\textsf{GAN}}_{\psi_d}(\mathcal{M}(f^{\textsf{GAN}}_{\psi_g}(\textbf{z})))$, $g_{\psi}^{\textsf{Gen}}(\textbf{z}) = h^{\textsf{GAN}}_{\psi_g}(\textbf{z})$ and $\textbf{z}\sim p_Z(\cdot)$.

\subsubsection{Limitations of generative model training using PPGC}
\begin{figure*}
    \centering
    \includegraphics[width=0.3\textwidth]{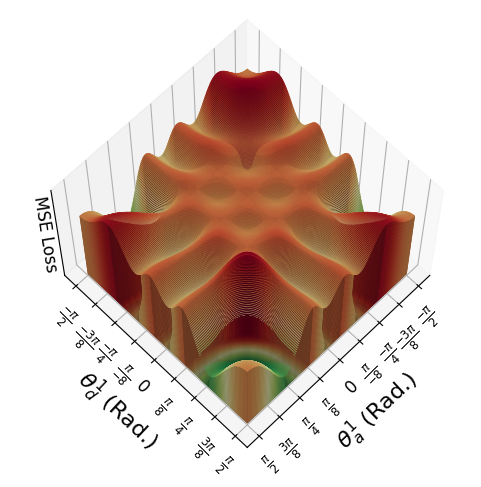}
    \includegraphics[width=0.3\textwidth]{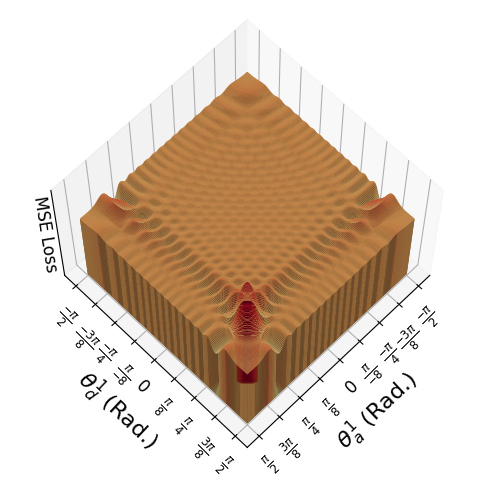}
    \includegraphics[width=0.3\textwidth]{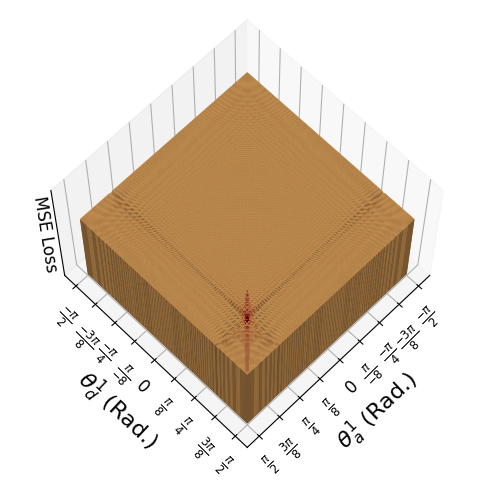}
    
    \includegraphics[width=0.3\textwidth]{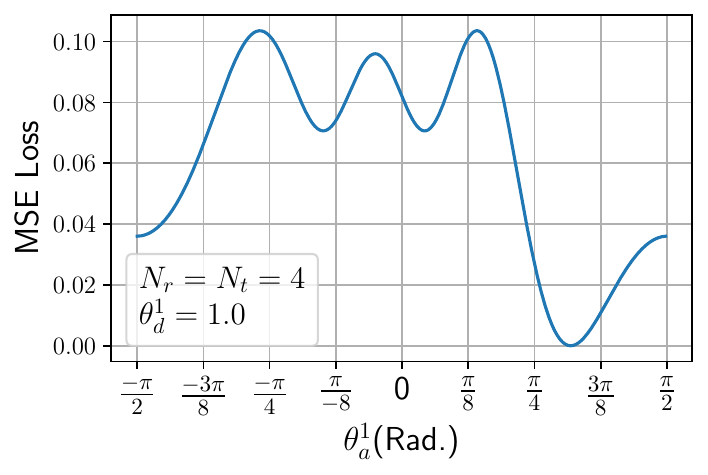}
    \includegraphics[width=0.3\textwidth]{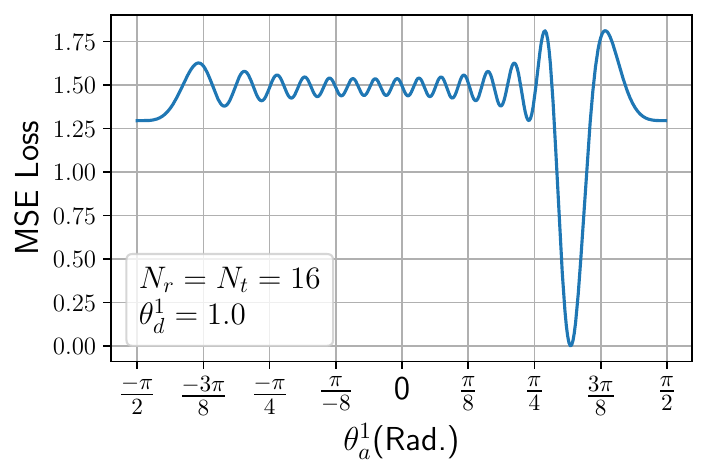}
    \includegraphics[width=0.3\textwidth]{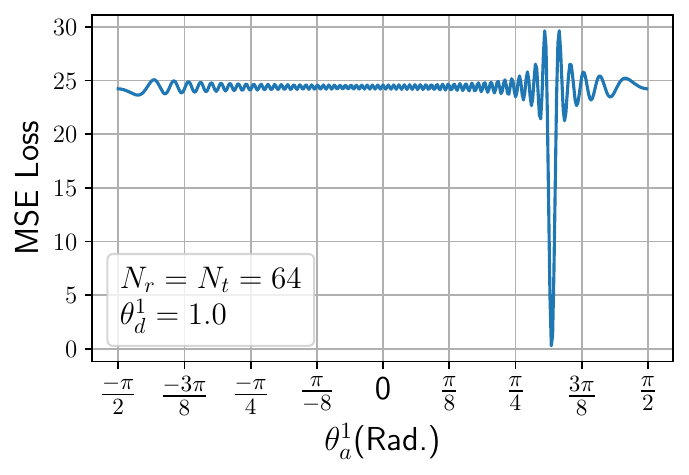}
    \caption{The loss surface as a function of $(\theta_a^1,\theta_d^1)$ in reference to a channel matrix with ground truth $\theta_a^1 = \theta_d^1 = 1.0$ Rad. using a PPGC model $\mathcal{M}$ for the azimuth plane using ULA antennas with $P=1$ and $N_r = N_t = \{4,16,64\}$ antennas respectively. The PPGC model $\mathcal{M}$ is extremely non-convex as a function of the parameters $\theta_a,\theta_d$ due to periodicity in the loss function arising from the formulation of the array response matrices. Avoiding the numerous local minima surrounding the global minimum poses a significant challenge given the nature of activation functions. Also, as the number of antennas $N_r,N_t$ increase, gradient flowing through the model is negligible at locations further away from the minima, where the loss surface is effectively flat. }
    \label{fig:nonconvex_surface}
\end{figure*}

In order to understand the limitations of training a generative model directly in conjunction with the PPGC model given by \eqref{eq:channel_sum}, we analyze the role of the UPA manifold, given by \eqref{eq:array_responses}. 
The presence of sinusoidal functions of the angular channel parameters ($\theta_{d}^{p},\theta_{a}^{p},\phi_{d}^{p},\phi_{a}^{p}$) in creating the array response matrices results in periodicities in the loss function landscape, with the resulting non-convexity increasing with the number of paths, as seen in Fig. \ref{fig:nonconvex_surface}.
These periodicities cannot be effectively approximated by the non-linear activation functions used in deep neural networks, leading to difficulties in training \cite{rectified_boltzman,guide_to_conv_aritmetic}. As a result, depending on the location of the optimizer in the parameter space, the gradient may not flow during backpropagation, resulting in the optimizer converging to unsuitable local minima. 
Additionally, as seen in Fig. \ref{fig:nonconvex_surface},
the convergence of the system to a suitable minimum is highly dependent on the number of antennas, $N_t, N_r$, used in the PPGC model. As $N_t, N_r$ increase, (i) the number of local minima also increases, and (ii) the optimality gap, that is, the difference between the loss values at the local and global minima, widens, causing convergence to any local minima to significantly impact the overall loss, as observed in Fig. \ref{fig:model_train}. Also, as observed in Fig. \ref{fig:nonconvex_surface}, the loss landscape flattens as parameter values move further from the global optimum, leading to smaller gradients. This prevents gradient-based optimization from converging to optimal parameter values, regardless of the strategies and momentum employed.


\subsection{Linearized Reformulation of the Physics Model}
\label{sec:pred_matrix}

To overcome the challenges posed by the PPGC model while maintaining the underlying model features for channel generation, we relax the overall cost function by reformulating the PPGC model $\mathcal{M}$ by first discretizing the range of values that $\theta_a^p,\theta_d^p,\phi_a^p,\phi_d^p$ can take and then expressing the channel generation process as a weighted sum of the resultant response matrices $\textbf{A}_r(\theta^p_a,\phi_a^p) \textbf{A}_t(\theta^p_d,\phi^p_d)^H$. Different from directly estimating the channel parameters, this method utilizes dictionary-based channel generation, where the loss function is a linear function of learnable weights, and is described next.
\begin{enumerate}
    \item Define the range of azimuth angles  $\theta_a^p,\theta_d^p$, given by $[\theta_{\min},\theta_{\max}]$, divided equally into $R_{\theta}$ intervals of width $\Delta_\theta = (\theta_{\max}-\theta_{\min})/R_{\theta}$. Similarly, the range of elevation angles  $\phi_a^p,\phi_d^p$, given by $[\phi_{\min},\phi_{\max}]$, is divided into $R_{\phi}$ intervals of width $\Delta_\phi = (\phi_{\max}-\phi_{\min})/R_{\phi}$. 
    \item Pre-compute the outer product between the array response matrices, $\textbf{A}_r(\theta^p_a,\phi^p_a) \textbf{A}_t(\theta^p_d,\phi^p_d)^H$, at the discretized angle values and store them in a dictionary $\textbf{D}$. 
\end{enumerate}
The dictionary $\textbf{D}$ has a total of $R_{\theta}^{2}R_{\phi}^2$ elements, henceforth referred to as the \emph{array response dictionary}. Each element of the dictionary, $\textbf{D}_{i,j,m,n}~ \forall~ i,j \in \{1,R_{\theta}\}, m,n \in \{1,R_{\phi}\}$,  represents a $N_r \times N_t$ matrix, and is given by

\begin{equation}
\label{eq:calc_array_dict}
    \textbf{D}_{i,j,m,n} = \textbf{A}_r(\theta_i,\phi_m) \textbf{A}_t(\theta_j,\phi_n)^{H},
\end{equation} 
where $\big\{\theta_k := \theta_{\min}+k(\Delta_\theta)|~  k \in \{i,j\}\big\}$ and $\big\{\phi_k := \phi_{\min}+k(\Delta_\phi)|~  k \in \{m,n\}\big\}$.
Thus, each element of the array response dictionary is the combination of antenna array responses at the transmitter and receiver for unique values of the angle of arrival and angle of departure. The angles associated with neighboring elements of the dictionary $\{\textbf{D}_{i,j,m,n},\textbf{D}_{i+1,j,m,n}\}$ or $\{\textbf{D}_{i,j,m,n},\textbf{D}_{i,j+1,m,n}\}$ differ by a value of $\Delta_\theta$ for the azimuth angle of arrival or the angle of departure respectively, while $\{\textbf{D}_{i,j,m,n},\textbf{D}_{i,j,m+1,n}\}$ or $\{\textbf{D}_{i,j,m,n},\textbf{D}_{i,j,m,n+1}\}$ differ by a value of $\Delta_\phi$ for the elevation angles of arrival or departure respectively. The relaxed PPGC model is now expressed as,

\begin{equation}
\label{eq:lincombi_channel}
    \textbf{H} = \sum_{i=1}^{R_{\theta}}\sum_{j=1}^{R_{\theta}}\sum_{m=1}^{R_{\phi}}\sum_{n=1}^{R_{\phi}}\textbf{W}_{i,j,m,n}\textbf{D}_{i,j,m,n},
\end{equation}
where, the channel generation is parametrized by the gain tensor $\textbf{W} \in \mathbb{R}^{R_{\theta} \times R_{\theta} \times R_{\phi} \times R_{\phi}}$, instead of the parameters $\textbf{s}$.
Each individual element of the gain tensor, given by $\mathbf{W}_{i,j,m,n}$ denotes the contribution of the path constructed from $\{\theta_{a}^{i},\theta_{d}^{j},\phi_{a}^{m},\phi_{a}^{n}\}$.The weight elements are expected to be non-zero for the paths corresponding to the ground-truth channel.
With this pipeline, we now model the output channel $\textbf{H}$ as a linear function of the gain tensor $\textbf{W}$, mitigating the issues arising from the non-convexity of the PPGC model $\mathcal{M}$ as seen in Sec.~\ref{sec:pred_params}. Thus, the generative model now predicts $\textbf{W}$ instead of $\textbf{s}$. Now, as the total number of paths $P$ is typically small, as mentioned in Sec.~\ref{sec:channel_model}, 
$\textbf{W}$ is expected to be sparse. 

\begin{figure*}
    \centering
    \includegraphics[trim={35cm 0 15cm 0},clip,width=0.98\textwidth]{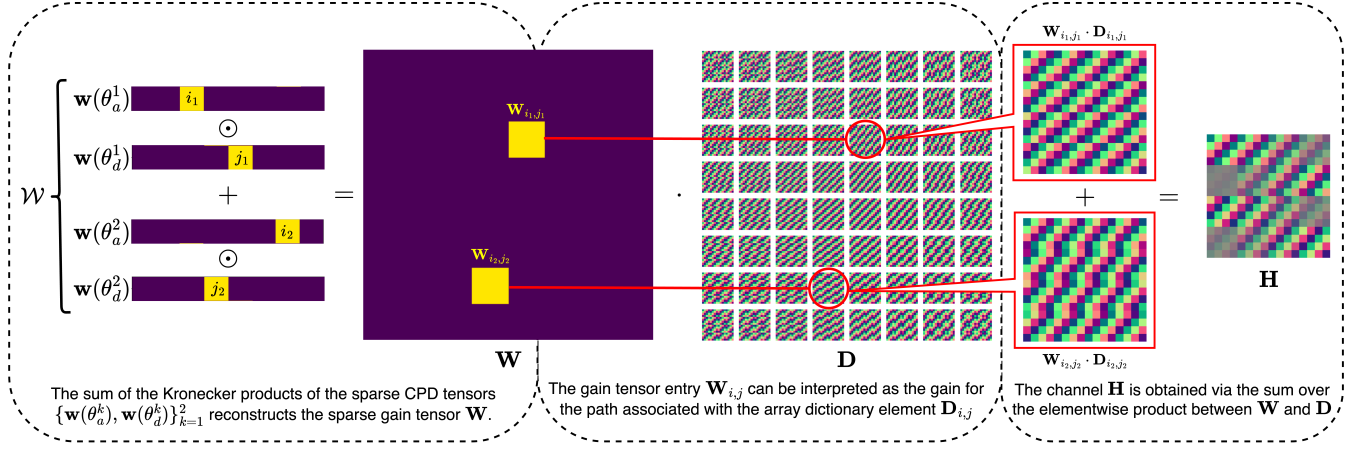}
    \caption{In order to visualize the channel construction process, we consider the 2-D projection of the azimuth angles $\theta_a,\theta_d$ for a channel with $2$ paths, with angles of elevation $\phi_a = \pi/3, \phi_d=\pi/3$ for both paths. The sparsity structure arising in the full gain tensor $\textbf{W}$ is also reflected in the sparse CPD vectors $\{\textbf{w}(\theta_a^k),\textbf{w}(\theta_d^k)\}_{k=1}^{2}$. The locations of the non-zero elements of the CPD vectors indicate the values of the associated angles. In this case, $(\textbf{w}(\theta_a^k),\textbf{w}(\theta_d^k))$ correspond to the azimuth angles of arrival and departure for the $k$-th path, respectively, relating to the associated elements in the array response dictionary $\textbf{D}$.} 
    \label{fig:cpd_algo}
\end{figure*}

\begin{remark} It is to be noted that (\ref{eq:channel_sum}) and (\ref{eq:lincombi_channel}) both produce valid channel matrices. For suitably high values of $R_{\theta},R_{\phi}$, any channel $\textbf{H}$ can be approximated by (\ref{eq:lincombi_channel}) using a suitable sparse $\textbf{W}$ with $P$ non-zero values. For such a tensor $\textbf{W}$, (\ref{eq:lincombi_channel}) is equivalent to (\ref{eq:channel_sum}), and each non-zero value $\textbf{W}_{i,j,m,n}$ will correspond to the gain $g_p$ of one of the $P$ paths. Thus, for this linearized representation of the problem, we can interpret the non-zero values of the gain tensor $\textbf{W}$ as the path gains, and the angles associated with the corresponding array response $\textbf{D}_{i,j,m,n}$ as the angles of arrival and departure of those paths. 
\end{remark}

\subsection{Decomposition of Parameter Tensor}
\label{sec:tensor_decomp}

\begin{figure}[h!]
    \centering
    \includegraphics[width=0.9\columnwidth]{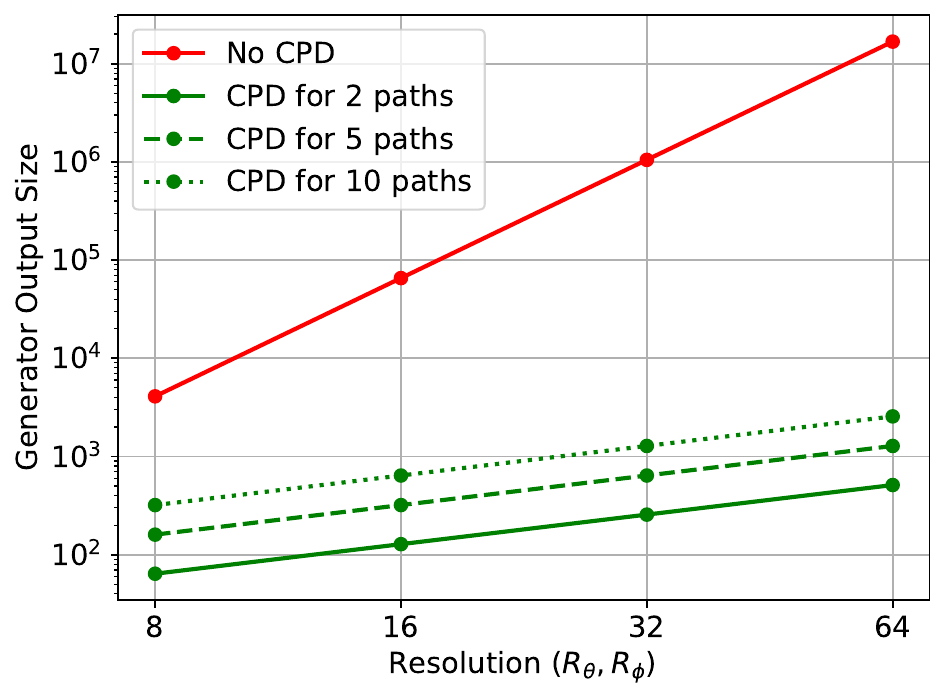}
    \caption{By predicting the Canonical Polyadic Decomposition (CPD) of the gain tensor $\textbf{W}$, the generator output size scales as $\mathcal{O}(RP)$, where $R=\max\{R_{\theta},R_{\phi}\}$ with the resolutions $R_{\theta},R_{\phi}$, reduced from $\mathcal{O}(R_{\theta}^{2} R^{2}_{\phi})$ without CPD. This allows the model to scale to an arbitrary number of paths $P$. 
    }
    \label{fig:output_size_scaling}
    \vspace{-4mm}
\end{figure}

The weight tensor $\textbf{W}$ as defined in Sec.\ref{sec:pred_matrix}, scales exponentially with the azimuth and elevation resolutions as $\mathcal{O}(R_{\theta}^{2} R_{\phi}^2)$. This results in significant practical challenges at higher resolutions of the linearized PPGC model, as seen in Fig. \ref{fig:output_size_scaling}. This issue is exacerbated by the introduction of new parameters, e.g. phase ($\eta$), in the channel model, which results in the size of $\textbf{W}$ scaling as $\mathcal{O}(R_{\theta}^2 R_{\phi}^2 R_{\eta})$, where $R_{\eta}$ is the resolution used to discretize the phase $\eta$. Thus, computing outer products becomes prohibitively expensive with at scale.


We mitigate this issue by modifying the generator $g_{\psi}$ to predict the canonical polyadic decomposition (CPD) \cite{cpd} of the gain tensor $\textbf{W}$, which is expressed as a set of CPD vectors $\mathcal{W} = \{\textbf{w}(\theta_{a}^{i}),\textbf{w}(\theta_{d}^{i}),\textbf{w}(\phi_{a}^{i}),\textbf{w}(\phi_{d}^{i})\}_{i=1}^{P}$, where $\textbf{w}(\theta_{a}^{i}),\textbf{w}(\theta_{d}^{i}) \in \mathbb{R}^{R_{\theta}} ~\forall ~ i \in 
\{1,P\} $ and $\textbf{w}(\phi_{a}^{i}),\textbf{w}(\phi_{d}^{i}) \in \mathbb{R}^{R_{\phi}} ~ \forall ~ i \in \{1,P\}$. We assume that the maximum number of paths $P$ constituting the channel matrix $\textbf{H}$ in (\ref{eq:channel_sum}) is known. We then obtain the full gain tensor $\textbf{W}$ by taking the sum of Kronecker products of the CPD vectors as

\begin{equation}
    \textbf{W} = \sum_{i=1}^P \textbf{w}(\theta_{a}^{i}) \otimes \textbf{w}(\theta_{d}^{i}) \otimes \textbf{w}(\phi_{a}^{i}) \otimes \textbf{w}(\phi_{d}^{i}).
    \label{eq:kronecker_product}
\end{equation}
By predicting the set of CPD vectors $\mathcal{W}$, the size of the generator output scales linearly with the resolution and the number of paths as $\mathcal{O}(RP)$, where $R = \max\{R_{\theta},R_{\phi}\}$, as opposed to $\mathcal{O}(R_{\theta}^2 R_{\phi}^2)$. This results in significant reductions in the output size of generator $g_{\psi}$, making the model tractable at larger resolutions, as shown in Fig. \ref{fig:output_size_scaling}. 

The choice of CPD as the tensor decomposition method is motivated by the fact that $\textbf{W}$ is a sparse tensor with $P$ non-zero entries, as shown in Remark 1. Therefore, $\textbf{W}$ has a rank of at most $P$, implying that $\textbf{W}$ can be decomposed to CPD vectors $\mathcal{W}$, and can be reconstructed via the sum of their Kronecker products as shown in (\ref{eq:kronecker_product}). The Kronecker product operation is differentiable, and is thus compatible with commonly used gradient descent based optimizers for the generator $g_{\psi}$. 

It should be noted that the CPD vectors $\mathcal{W}$ retain the interpretability characteristics desired from the generator output. As described in Sec. \ref{sec:pred_matrix}, the azimuth and elevation angles of arrival and departure are captured by the locations of non-zero entries in $\textbf{W}$, where a single non-zero entry $\textbf{W}_{i,j,m,n}$ corresponds to the angles for a single path. The CPD of the sparse gain tensor $\textbf{W}$ results in row-sparse CPD vectors $\mathcal{W}$, as shown in Fig. \ref{fig:cpd_algo}, where each CPD vector corresponds to one of the angles, where the order of angles depends on the specific construction of the array response dictionary $\textbf{D}$. For example, if $\textbf{D}$ is given by (\ref{eq:calc_array_dict}), the order of the CPD vectors must follow (\ref{eq:kronecker_product}). Now, the location of the non-zero entries of each CPD vector indicates the value of the corresponding angle. 

\begin{remark}
 In contrast to purely generative models, the computational complexity of our physics-based generative pipeline is independent of the size of the output channel matrix, owing to the reformulated dictionary $\textbf{D}$. As such, the complexity of our method can be tailored to user preference, as it scales linearly with the resolution $R$ and the maximum number of paths $P$, both of which are user-defined. For practical implementations, the memory footprint of $\textbf{D}$ can be reduced by decomposing it into two tensors as $\textbf{D} = \textbf{D}^r(\textbf{D}^t)^H$, where $\textbf{D}^r = [\textbf{A}_r(\theta_i,\phi_m)]_{i=0,m=0}^{i=R_{\theta},m=R_{\phi}}$ and $\textbf{D}^t = [\textbf{A}_t(\theta_j,\phi_n)]_{j=0,n=0}^{j=R_{\theta},n=R_{\phi}}$, similar to (\ref{eq:calc_array_dict}). By storing $\textbf{D}^r, \textbf{D}^t$ instead of $\textbf{D}$, we reduce the memory footprint of $\textbf{D}$ from $\mathcal{O}(R^4)$ to $\mathcal{O}(2R^2)$.
\end{remark}

\section{Integration of Reformulated Channel Model with Learning Algorithms}

\subsection{Generative Model for Synthetic Channel Generation}

\begin{figure*}
    \centering
    \includegraphics[trim={0 0 0 0},clip, width=0.98\textwidth]{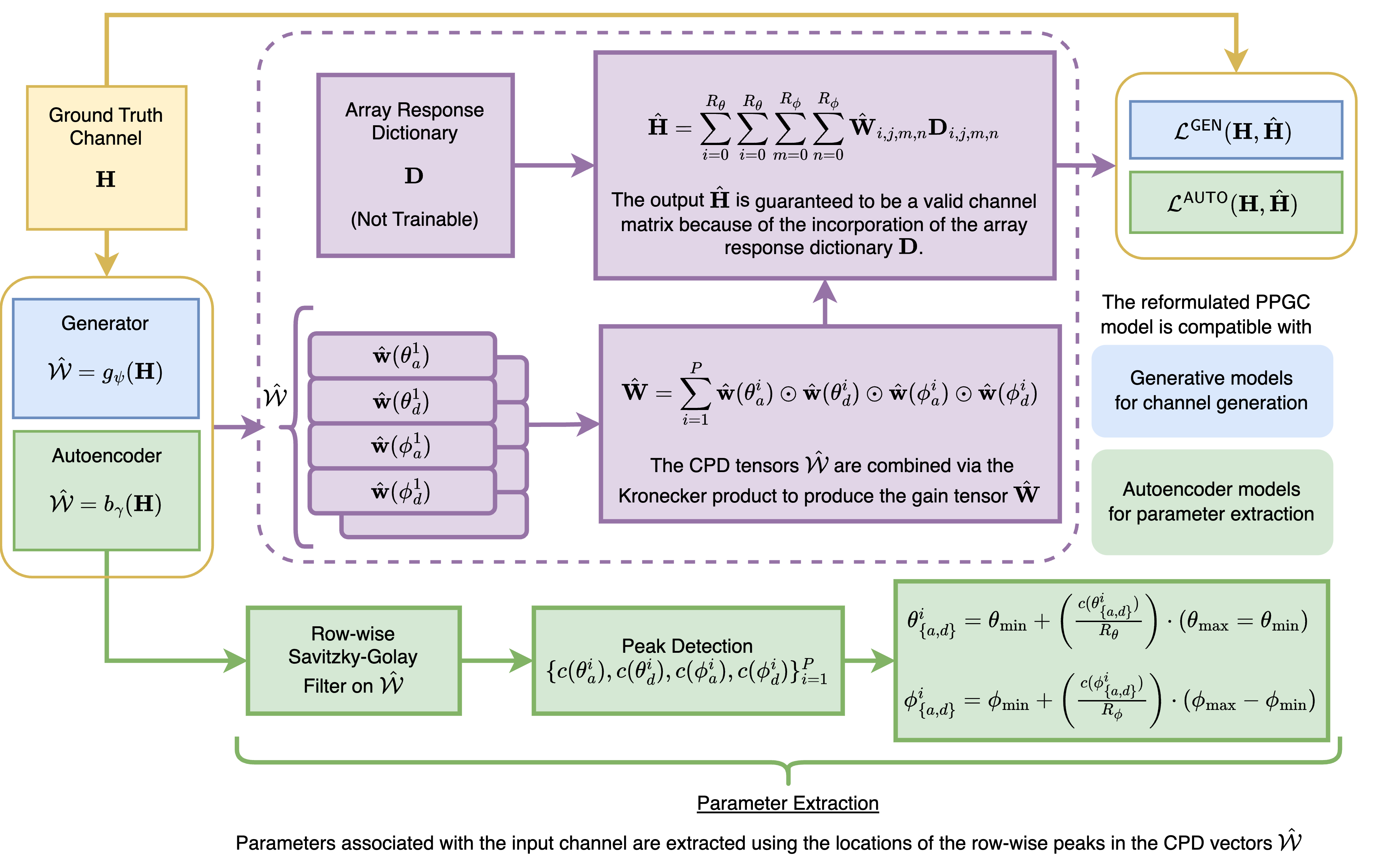}
    \caption{We reformulate the PPGC model by defining a discretized array response dictionary $\textbf{D}$ and use the generative model to produce CPD vectors $\hat{\mathcal{W}}$, which are then combined using the Kronecker Product to produce the gain tensor $\hat{\textbf{W}}$. The elementwise multiplication of $\hat{\textbf{W}}$ and $\textbf{D}$ mimics the PPGC model process. The reformulated PPGC model can be used with generative models to produce valid channel matrices that belong to the distribution of ground truth channels (blue blocks), as well as with autoencoders to explicitly extract the values of azimuth and elevation angles associated with input channels through a separate parameter extraction module which is independent of autoencoder training (green blocks). This reformulation allows the flow of gradient through the models, enabling it to converge to more suitable optima.}
    \label{fig:limmodel_train}
    \vspace{-3mm}
\end{figure*}

The generative pipeline incorporating the reformulated PPGC model can be used to generate new samples of channel matrices that are guaranteed to be valid in terms of the chosen channel model. In this section, we describe the processes for training and sampling these generative models.

\begin{algorithm}[tb]
\caption{Generation of Channel Data using Reformulated PPGC Model}
\label{algo:lin_model}

\underline{\textbf{\textit{\textbf{Training}}} ($g_{\psi}^{\textsf{Train}}$)} \\
\textbf{Given:} Dataset of valid channels $\mathcal{D}$, generator parameters $\psi$, Range of angles $[\theta_{\min},\theta_{\max},\phi_{\min},\phi_{\max}]$, Resolution $R_{\theta},R_{\phi}$.
\begin{algorithmic}[1]
    \STATE Calculate the array response dictionary $\textbf{D}$ using (\ref{eq:calc_array_dict}) for all $i,j \leq R_{\theta}, m,n \leq R_{\phi}$.
    \STATE Sample a channel matrix $\textbf{H}$ from the dataset $\mathcal{D}$.
    \STATE Produce a set of CPD vectors $\hat{\mathcal{W}}$ from $\textbf{H}$ via (\ref{eq:predict_cpd_vectors}).
    \STATE Obtain gain tensor $\hat{\textbf{W}}$ from CPD vectors via (\ref{eq:kronecker_product}).
    \STATE Obtain the predicted channel matrix $\hat{\textbf{H}}$ from $\hat{\textbf{W}}$ using (\ref{eq:vae_sum_weight}). 
    \STATE Calculate the loss using (\ref{eq:vae_loss}) and update parameters $\psi$.
\end{algorithmic}
\textbf{Output: } Trained generator model parameters $\psi$.

\underline{\textbf{\textit{\textbf{Generation}}} ($g_{\psi}^{\textsf{Gen}}$)} \\
\textbf{Given:} Sample vector $\widetilde{\textbf{z}} \in \mathbb{R}^{Z}$ from unit Gaussian distribution $\mathcal{N}(0,\textbf{I})$, trained generator $g_\psi$.
\begin{algorithmic}[1]
    \STATE Generate a set of CPD vectors $\mathcal{W}$ from $\widetilde{\textbf{z}}$ via (\ref{eq:gen_cpd_vectors}).
    \STATE Obtain gain tensor $\widetilde{\textbf{W}}$ from CPD vectors via (\ref{eq:kronecker_product}).
    \STATE Obtain generated channel matrix $\widetilde{\textbf{H}}$ from $\widetilde{\textbf{W}}$ using (\ref{eq:gen_channel}).
\end{algorithmic}
\textbf{Output: } Generated channel matrix $\widetilde{\textbf{H}}$.
\end{algorithm}

\emph{Training: }
To train the generative model to predict the CPD vectors $\mathcal{W}$, we change the pipeline described in Sec. \ref{sec:pred_params} in the following ways. The generator $g_{\psi}^{\textsf{Train}}:\mathbb{R}^{2 \times N_t \times N_r}\rightarrow\mathbb{R}^{4RP}$ now predicts the CPD vectors $\hat{\mathcal{W}}$, as defined in Sec. \ref{sec:tensor_decomp}, as

\begin{equation}
    \hat{\mathcal{W}} = g_{\psi}^{\textsf{Train}}(\textbf{H}).
    \label{eq:predict_cpd_vectors}    
\end{equation}
The CPD vectors $\hat{\mathcal{W}}$ are then combined as the sum of their Kronecker products to produce the predicted gain tensor $\hat{\textbf{W}}$, as defined in (\ref{eq:kronecker_product}). $\hat{\textbf{W}}$ is then used to generate the predicted channel $\hat{\textbf{H}}$ by multiplying it with the array response dictionary as
\begin{equation}
\label{eq:vae_sum_weight}
    \hat{\textbf{H}} = \sum_{i=1}^{R_{\theta}}\sum_{j=1}^{R_{\theta}}\sum_{m=1}^{R_{\phi}}\sum_{n=1}^{R_{\phi}} \hat{\textbf{W}}_{i,j,m,n} \textbf{D}_{i,j,m,n}.
\end{equation}
Now, we propose a modified system loss which operates on the predicted channel $\hat{\textbf{H}}$ and the input $\textbf{H}$, that also accounts for the sparsity in the CPD vectors $\mathcal{W}$. This is given by
\begin{equation}
\label{eq:vae_loss}
    \mathcal{L} = \mathcal{L}^{\textsf{GEN}}(\textbf{H},\hat{\textbf{H}}) + \tau \left( \sum_{\hat{\textbf{w}} \in \hat{\mathcal{W}}} ||\hat{\textbf{w}}||_1\right),
\end{equation}
where the first term, $\mathcal{L}^{\textsf{GEN}}$, is the architecture-specific loss function defined for generator models as described in Sec. \ref{sec:pred_params}. The last term is a regularization term that penalizes $\ell_{1}$-norm of the CPD vectors $\hat{\mathcal{W}}$, encouraging sparsity in the CPD vectors $\hat{\mathcal{W}}$, thereby causing the gain tensor $\hat{\textbf{W}}$ to also be sparse. Hence, the overall loss function improves the fidelity of the generator outputs through the first term, while ensuring that $\hat{\textbf{W}}$ does not select a large combination of paths that do not correspond to realistic scenarios,
enabling the model to estimate sparse CPD vectors from the input channel matrix.

\emph{Inference and Sampling : }
During the inference phase, we only use the generator $g_{\psi}$ in generation mode to produce new channel matrices. First, we sample a vector $\widetilde{\textbf{z}} \sim \mathcal{N}(0,\textbf{I})$ and use the generative process to produce CPD vectors as

\begin{equation}
\label{eq:gen_cpd_vectors}
    \widetilde{\mathcal{W}} = g_{\psi}^{\textsf{Gen}}(\widetilde{\textbf{z}}).
\end{equation}

The generated CPD vectors $\widetilde{\mathcal{W}}$ are then combined as the sum of their Kronecker products to produce a new example of the gain tensor $\widetilde{\textbf{W}}$, as defined in (\ref{eq:kronecker_product}).
$\widetilde{\textbf{W}}$ is then used to generate a synthetic channel, using the array response dictionary, as
\begin{equation}
\label{eq:gen_channel}
    \widetilde{\textbf{H}} = \sum_{i=1}^{R_{\theta}}\sum_{j=1}^{R_{\theta}}\sum_{m=1}^{R_{\phi}}\sum_{n=1}^{R_{\phi}} \widetilde{\textbf{W}}_{i,j,m,n} \textbf{D}_{i,j,m,n}.
\end{equation}
The generated channel $\widetilde{\textbf{H}}$ is guaranteed to be a valid channel matrix as the mapping from the parameter space to the channel space is done using the array response dictionary $\textbf{D}$, which is based on a verified PPGC model. $\widetilde{\textbf{H}}$ is also expected to be from the distribution of interest, as the first term in (\ref{eq:vae_loss}) ensures that the generator learns to map samples from the unit Gaussian distribution to a set of parameters that belong to the target distribution. 
A summary of the training and generative processes for the linearized representation is given in Algo. \ref{algo:lin_model}.

\begin{algorithm}[tb]
\caption{PPGC Parameter Extraction using Autoencoders}
\label{algo:cpd_autoencoder_algo}

\underline{\textit{\textbf{Training}}} :\\
\textbf{Given:} Dataset of valid channels $\mathcal{D}$, autoencoder parameters $b_{\gamma}$, Range of angles $[\theta_{\min},\theta_{\max},\phi_{\min},\phi_{\max}]$, Resolution $R_{\theta},R_{\phi}$.
\begin{algorithmic}[1]
    \STATE Calculate the array response dictionary $\textbf{D}$ using (\ref{eq:calc_array_dict}) for all $i,j \leq R_{\theta}, m,n \leq R_{\phi}$.
    \STATE Sample a channel matrix $\textbf{H}$ from the dataset $\mathcal{D}$.
    \STATE Produce a set of CPD vectors $\hat{\mathcal{W}}$ from $\textbf{H}$ via (\ref{eq:cpd_autoencoder}).
    \STATE Obtain gain tensor $\hat{\textbf{W}}$ from CPD vectors via (\ref{eq:kronecker_product}).
    \STATE Obtain the predicted channel matrix $\hat{\textbf{H}}$ from $\hat{\textbf{W}}$ using (\ref{eq:vae_sum_weight}). 
    \STATE Calculate the loss using (\ref{eq:cpd_autoencoder_loss}) and update $b_{\gamma}$.
\end{algorithmic}
\textbf{Output:} Trained autoencoder $b_{\gamma}$

\underline{\textit{\textbf{Parameter Extraction}}} :\\
\textbf{Given:} Trained autoencoder $b_{\gamma}$, channel matrix $\widetilde{\textbf{H}}$.
\begin{algorithmic}[1]
    \STATE Generate a set of CPD vectors $\widetilde{\mathcal{W}}$ from $\widetilde{\textbf{H}}$ via (\ref{eq:cpd_autoencoder}).
    \STATE Apply Savitzky-Golay filter to rows of $\widetilde{\textbf{w}} \in \widetilde{\mathcal{W}}$.
    \STATE Obtain locations of peaks $\{c(\theta_a^i),c(\theta_d^i),c(\phi_a^i),c(\phi_d^i)\}_{i=1}^p$.
    \STATE Calculate angle values $\{\hat{\theta_a^i},\hat{\theta_d^i},\hat{\phi]^i},\hat{\phi_d^i}\}_{i=1}^P$ using (\ref{eq:estimate_angles}).
\end{algorithmic}
\textbf{Output: } Estimated parameter values $\{\hat{\theta_a^i},\hat{\theta_d^i},\hat{\phi]^i},\hat{\phi_d^i}\}_{i=1}^P$
\end{algorithm}

\subsection{Autoencoder Model to Extract Channel Parameters}
\label{sec:parameter_estimation}

In addition to channel generation, which is the primary function of the proposed model, its ability to recover CPD vectors can be extended to explicitly extract channel parameters $\textbf{s}$ from the associated channel matrix $\textbf{H}$.
We describe this process next. 

We consider an autoencoder $b_{\gamma}$ parametrized by $\gamma$, which takes a channel matrix $\textbf{H}$ as input and produces the CPD vectors $\hat{\mathcal{W}}$ as output, as 
\begin{equation}
    \hat{\mathcal{W}} = b_{\gamma}(\textbf{H}).
    \label{eq:cpd_autoencoder}
\end{equation}
The CPD vectors $\hat{\mathcal{W}}$ are then combined as the sum of their Kronecker products to obtain the predicted gain tensor $\hat{\textbf{W}}$, as defined in (\ref{eq:kronecker_product}). The gain tensor $\hat{\textbf{W}}$ is then used to produce the predicted channel $\hat{\textbf{H}}$ using (\ref{eq:vae_sum_weight}). The autoencoder is trained using the Mean Square Error (MSE) loss with an L1 regularization term as follows
\begin{equation}
    \mathcal{L}^{\textsf{AUTO}} = ||\hat{\textbf{H}}-\textbf{H}||_2^2 + \tau \left( \sum_{\hat{\textbf{w}} \in \hat{\mathcal{W}}} ||\hat{\textbf{w}}||_1\right).
    \label{eq:cpd_autoencoder_loss}
\end{equation}
Here, similar to (\ref{eq:vae_loss}), the first term minimizes the accuracy of the reconstructed channel $\hat{\textbf{H}}$, and the second term ensures that a large combination of paths are not selected.

\emph{Parameter Extraction : } In order to extract the parameters associated with the predicted channel $\textbf{H}$, we perform 1-D peak detection across the rows of each CPD vector $\hat{\textbf{w}} \in \hat{\mathcal{W}}$ by first applying the Savitzky-Golay filter \cite{savgol} to each row of the CPD vectors to remove noise, and then identifying the peaks above a threshold value. The peak detection algorithm extracts the locations of peaks in each row of the CPD vectors, given by $\{c(\theta_a^i),c(\theta_d^i),c(\phi_a^i),c(\phi_d^i)\}_{i=1}^P$, which correspond to the associated values of angles $\{\hat{\theta_a^i},\hat{\theta_d^i},\hat{\phi_a^i},\hat{\phi_d^i}\}_{i=1}^P$, as described in Sec. \ref{sec:tensor_decomp}. We then calculate the values of the associated angles of azimuth and elevation as follows
\begin{align}
    \hat{\theta_a^i} = \theta_{\min} + \left(\frac{c(\theta_a^i)}{R_{\theta}}\right)\cdot(\theta_{\max}-\theta_{\min}),\nonumber\\
    \hat{\phi_a^i} = \phi_{\min} + \left(\frac{c(\phi_a^i)}{R_{\phi}}\right)\cdot(\phi_{\max}-\phi_{\min}),
\label{eq:estimate_angles}
\end{align}
with corresponding equations for the angles of departure $\{\theta_d^i,\phi_d^i\}_{i=1}^P$, where $i\in\{1,\dots,P\}$ denote the path index. This process is independent of the training process of the autoencoder $b_{\gamma}$. The granularity of estimation depends upon the resolution $R_{\theta},R_{\phi}$, and range $[\theta_{\min},\theta_{\max},\phi_{\min},\phi_{\max}]$ of the angles of azimuth and elevation. The estimation granularities of azimuth and elevation angles are given by $\Delta_{\theta} = (\theta_{\max}-\theta_{\min})/R_{\theta}$ and $\Delta_{\phi} = (\phi_{\max}-\phi_{\min})/R_{\phi}$ respectively.
The autoencoder training and parameter extraction processes are illustrated in Fig. \ref{fig:limmodel_train}, and described in Algo. \ref{algo:cpd_autoencoder_algo}.

\begingroup
\setlength{\tabcolsep}{1.4pt} 
\renewcommand{\arraystretch}{1.1} 
\begin{table*}[t]
    \small
    \centering
    \begin{tabular}{|c||c|c|c|c|c|c|c|c|c|c|}
    \hline
       & \multicolumn{5}{|c|}{2-Wasserstein Distance}  & \multicolumn{5}{|c|}{MMD}\\
      \hline
      Dataset   & Ours & CGAN & UNet & CVAE & Glow & Ours & CGAN & UNet & CVAE & Glow\\
      \hline
      Boston  & \textbf{0.152}\footnotesize{$\pm$0.006} & 0.357\footnotesize{$\pm$0.002} & 0.853\footnotesize{$\pm$0.008}& 2.084\footnotesize{$\pm$0.043} & 1.384\footnotesize{$\pm$0.051} & \textbf{0.007}\footnotesize{$\pm$0.001} & 0.058\footnotesize{$\pm$0.005} & 1.753\footnotesize{$\pm$0.039} & 0.367\footnotesize{$\pm$0.008} & 1.301\footnotesize{$\pm$0.091} \\
       ASU    & \textbf{0.198}\footnotesize{$\pm$0.011} & 0.544\footnotesize{$\pm$0.003} & 0.952\footnotesize{$\pm$0.016} & 3.162\footnotesize{$\pm$0.12} & 0.938\footnotesize{$\pm$0.095} & \textbf{0.010}\footnotesize{$\pm$0.002} & 0.055\footnotesize{$\pm$0.005} & 0.441\footnotesize{$\pm$0.091} & 3.576\footnotesize{$\pm$0.220} & 0.275\footnotesize{$\pm$0.096} \\
       Indoor & \textbf{0.046}\footnotesize{$\pm$0.003} & 0.294\footnotesize{$\pm$0.008} & 1.349\footnotesize{$\pm$0.012} & 4.854\footnotesize{$\pm$0.11} & 0.569\footnotesize{$\pm$0.072} & \textbf{0.012}\footnotesize{$\pm$0.002} & 0.084\footnotesize{$\pm$0.018} & 0.332\footnotesize{$\pm$0.005} & 3.823\footnotesize{$\pm$0.120} & 0.142\footnotesize{$\pm$0.012}  \\
       Outdoor& \textbf{0.091}\footnotesize{$\pm$0.002} & 0.541\footnotesize{$\pm$0.006} & 0.877\footnotesize{$\pm$0.003} & 3.552\footnotesize{$\pm$0.11} & 0.596\footnotesize{$\pm$0.063} & \textbf{0.041}\footnotesize{$\pm$0.006} & 0.18\footnotesize{$\pm$0.018} & 2.246\footnotesize{$\pm$0.662} & 1.140\footnotesize{$\pm$ 0.121} & 0.188\footnotesize{$\pm$0.036}  \\
       \hline
    \end{tabular}
    \caption{The distribution of channels modeled by our method is closer to the ground truth channel distribution compared to baselines in terms of both 2-Wasserstein distance and Maximum Mean Discrepancy (MMD).}
    \label{tab:channel_dist}
\end{table*}
\endgroup

\begingroup
\setlength{\tabcolsep}{1.5pt} 
\renewcommand{\arraystretch}{1.3} 
\begin{table*}[t]
    \small
    \centering
    \begin{tabular}{|c||c c c c c|c c c c c|c c c c c| c c c c c|}
    \hline
       \multirow{2}{2em}{Train on} & \multicolumn{5}{|c|}{Testing on $\mathcal{D}^R_{10}$} & \multicolumn{5}{|c|}{Testing on $\mathcal{D}^R_{11}$} & \multicolumn{5}{|c|}{Testing on $\mathcal{D}^G_{10}$} & \multicolumn{5}{|c|}{Testing on $\mathcal{D}^G_{11}$} \\
        \cline{2-21}
    & Ours & CGAN & UNet & CVAE & Glow & Ours & CGAN & UNet & CVAE & Glow &  Ours & CGAN & UNet & CVAE & Glow &  Ours & CGAN & UNet & CVAE & Glow   \\
    \hline
    
    $\mathcal{D}^R_{10}$ & 0.05 & 0.05 & 0.05 &0.05 &0.05 & \lightgray{1.05} & \lightgray{1.05} & \lightgray{1.05} & \lightgray{1.05}& \lightgray{1.05} & \textbf{0.08} & 0.14  & 0.44 & 0.67 & 0.292 & \lightgray{1.07} & \lightgray{1.01} & \lightgray{1.0} & \lightgray{1.75} & \lightgray{1.315} \\
    \hline
    
    $\mathcal{D}^R_{11}$ & \lightgray{0.87} & \lightgray{0.87} & \lightgray{0.87} & \lightgray{0.87}& \lightgray{0.87} & 0.07 & 0.07 & 0.07 & 0.07& 0.07 & \lightgray{0.88} & \lightgray{1.15} & \lightgray{0.91} & \lightgray{1.14}& \lightgray{1.522} & \textbf{0.15} & 0.18 &  0.49 & 1.12 & 0.65\\
    \hline
    
    $\mathcal{D}^G_{10}$ & \textbf{0.07} & 0.37 & 0.44 & 0.8 & 0.842 & \lightgray{1.41} & \lightgray{1.11} & \lightgray{1.62} & \lightgray{1.15} & \lightgray{2.523} & 0.06 & 0.09 & 0.1 &0.19& 0.208& \lightgray{1.06} & \lightgray{1.11} & \lightgray{1.16} & \lightgray{2.24}& \lightgray{1.676} \\
    \hline
    $\mathcal{D}^G_{11}$ & \lightgray{1.27} & \lightgray{1.51} & \lightgray{1.2} & \lightgray{1.03} & \lightgray{1.073} & \textbf{0.12} & 0.33 & 0.21 & 0.75& 0.475 & \lightgray{1.29} & \lightgray{1.85} & \lightgray{0.94} & \lightgray{0.99} & \lightgray{1.856} & 0.02 & 0.03 & 0.08 & 0.11 & 0.187\\
    \hline

    \end{tabular}
    \caption{Cross evaluation of downstream compression tasks for different combinations of real and generated datasets by comparing NMSE loss. When compression models are trained on real data and tested on generated data (Rows $1$,$2$) and vice versa (Rows $3$,$4$), our method records lower NMSE for corresponding real-generated dataset pairs, indicating that the data generated by our method is more similar to the true channel data from the target ground truth distribution.}
    \label{tab:cross_eval}
    \vspace{-3mm}
    
\end{table*}
\endgroup

\section{Experiments}
\label{sec:experiments}

In this section, we analyze the performance of our method on wireless channel datasets corresponding to real-life scenarios and compare our method against prior baselines. We show the efficacy of our method in capturing the underlying parameter distributions, its high fidelity in generating channels, and its efficacy in downstream channel compression tasks.

For our PPGC model, we consider a system with $N_t = N_r = 16$ transmit and receive antennas respectively, with $4$ antennas in each row. We consider six datasets, five of which are generated by the DeepMIMO framework \cite{deepmimo} that utilizes 3D ray tracing for generating channel datasets in different settings. Datasets corresponding to the following scenarios are used; (i) Two base stations in an outdoor intersection of two streets with blocking and reflecting surfaces. The channels corresponding to  base station 10 (BS10) and 11 (BS11) have been considered; (ii) An indoor conference room (Indoor); (iii) A section of downtown Boston, Massachusetts, USA, generated using the 5G model developed by RemCom \cite{remcom} (Boston) and (iv) A section of the Arizona State University campus in Tempe, Arizona, USA (ASU). The sixth dataset is a user-defined dataset of size $20,000$ produced by sampling known distributions of parameters to use with the PPGC model $\mathcal{M}$ with $5$ paths to produce a dataset (User Defined). The distribution over the parameter space for each path is given in Table \ref{tab:path_dists}. For the Boston and ASU datasets, we consider $2$ paths and for the Indoor, BS10 and BS11 datasets, we consider $3$ paths. It should be noted that for datasets simulated using the DeepMIMO framework, the number of paths are limited by the number of datapoints available in each scenario. All datasets are split $80/20$ for training and testing. 

\begingroup
\setlength{\tabcolsep}{2pt} 
\renewcommand{\arraystretch}{1.1} 
\begin{table}[h!]
    \small
    \centering
    \begin{tabular}{|c|c|c|c|c|}
    \hline
       $p$  & $\theta_a^p$ (Radians) & $\theta_d^p$ (Radians) & $\phi_a^p$ (Radians) & $\phi_d^p$ (Radians)\\
       \hline
        1 & $\mathcal{U}(0.55,0.85)$ & $\mathcal{U}(0.55,0.85)$ & $\mathcal{U}(1.15,1.45)$ & $\mathcal{U}(1.47,1.67)$ \\
        2 & $\mathcal{U}(-0.85,-0.25)$ & $\mathcal{U}(-0.85,-0.25)$ & $\mathcal{U}(1.47,1.67)$ & $\mathcal{U}(1.6,1.8)$\\
        3 & $\mathcal{U}(0.45,0.75)$ & $\mathcal{U}(-0.75,-0.45)$ &$\mathcal{U}(1.7,1.9)$&$\mathcal{U}(1.42,0.72)$\\
        4 & $\mathcal{U}(-0.1,0.3)$ & $\mathcal{U}(0.4,0.8)$ &$\mathcal{U}(1.15,1.35)$&$\mathcal{U}(1.55,1.85)$\\
        5 & $\mathcal{U}(-1.0,-0.6)$ & $\mathcal{U}(0.1,0.5)$ &$\mathcal{U}(1.55,1.85)$&$\mathcal{U}(1.15,1.35)$\\
        \hline
    \end{tabular}
    \caption{Distributions of channel parameters for each path used to generate the User Defined dataset.}
    \label{tab:path_dists}
    \vspace{-5mm}
\end{table}
\endgroup

We use a VAE as the generative model $g_{\psi}$, with the encoder and decoder given by $(f^{\textsf{VAE}}_{\psi_e},f^{\textsf{VAE}}_{\psi_d})$ respectively. The VAE encoder $f^{\textsf{VAE}}_{\psi_e}$ consists of four convolutional layers, with kernel size $2$ and stride $1$, followed by two linear layers with output sizes $1024$ and $512$. The layers used for sampling the latent distribution are linear layers with output size $64$. The VAE decoder $f^{\textsf{VAE}}_{\psi_d}$ consists of five linear layers with output sizes $256,1024,2048,4096$ and $2P(R_{\theta}+R_{\phi})$ respectively.
We compare our model against  ChannelGAN (CGAN)\cite{channelgan}, the DUNet diffusion model (UNet) \cite{diffusion_models_for_channels}, VAE version of CSINet \cite{csinet} and the GLOW generative flow model (Glow)\cite{glow}. Models are trained for $300$ epochs with a batch size of $256$. For our model, we use resolution $R_{\theta} = 32$ and $R_{\phi} = 16$.

\begin{figure*}[t]
    \centering
    \includegraphics[trim={0 0 0 0},clip, width=0.31\textwidth]{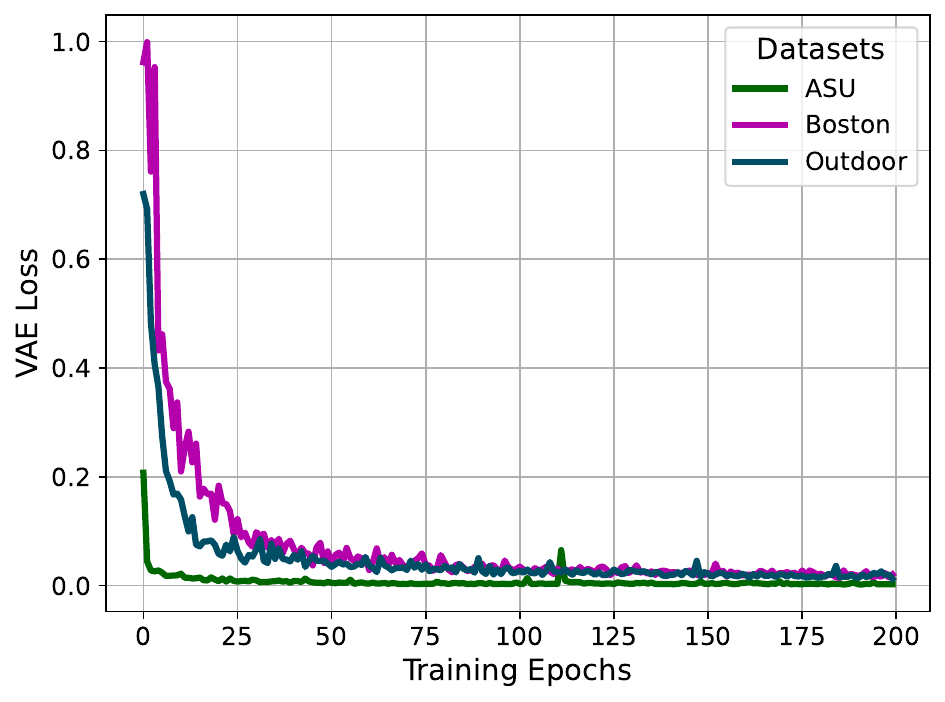}
    \includegraphics[trim={0 0 0 0},clip, width=0.31\textwidth]{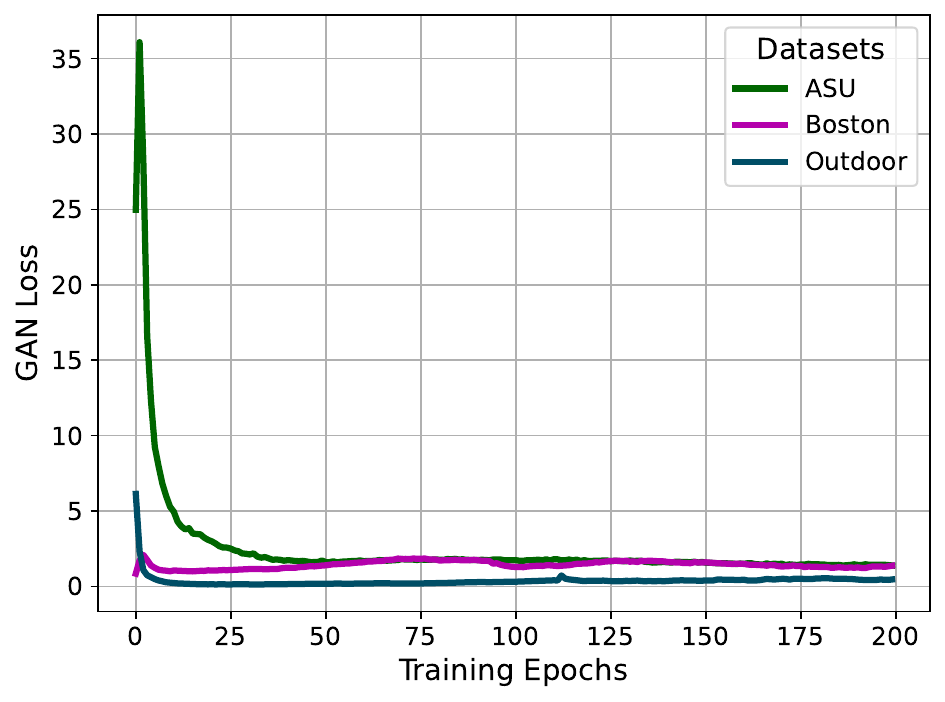}
    \includegraphics[trim={0 0 0 0},clip, width=0.34\textwidth]{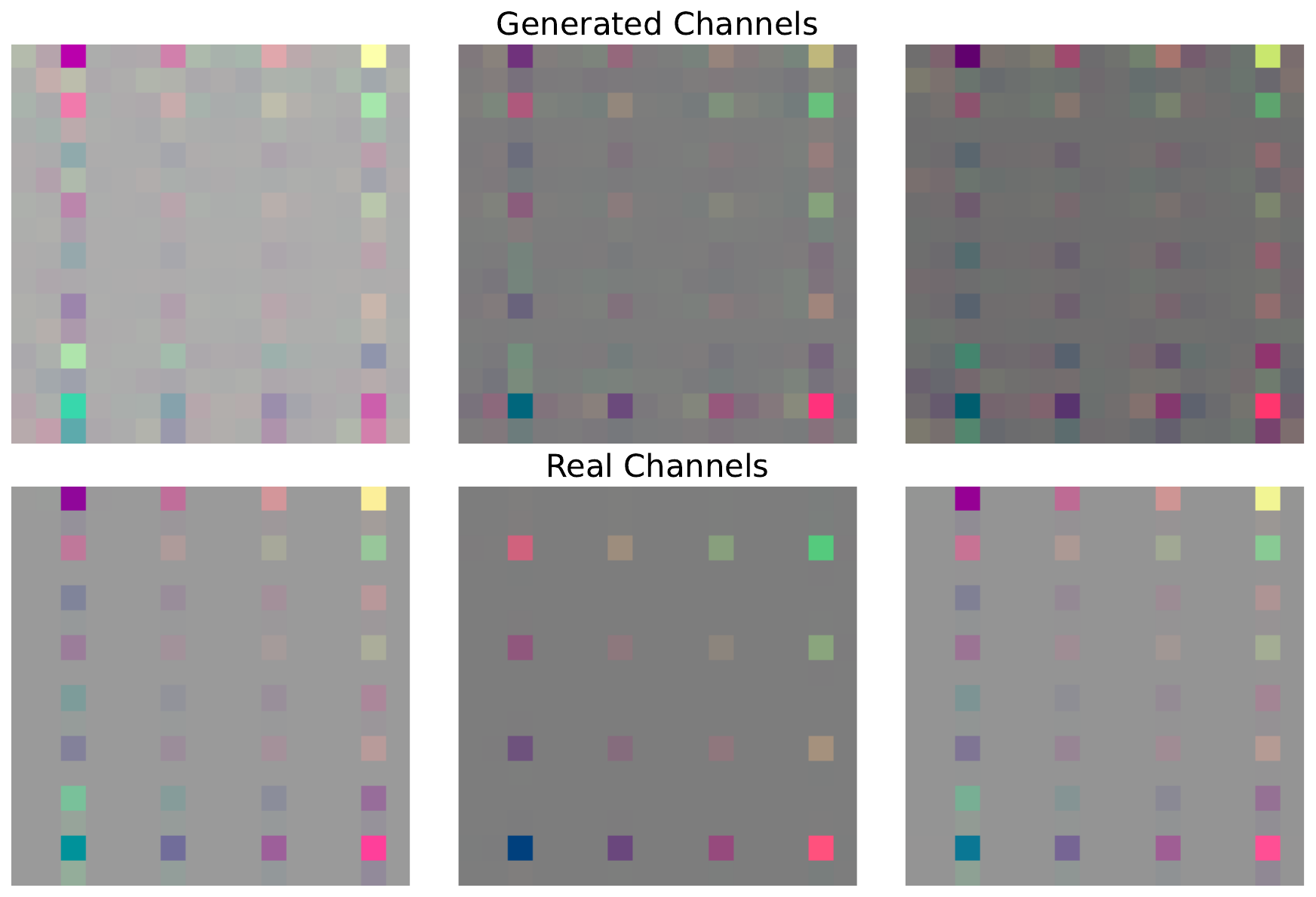}
    \caption{Our method is compatible with the VAE (left) and GAN (center) architectures, with generator models converging to suitable optima. The inclusion of the array response dictionary $\textbf{D}$ in the training pipeline guarantees valid channel matrices at the generator output that are similar to the ground truth channels, thus capturing the underlying parameter distributions (right).}
    \label{fig:exp_compat}
    \vspace{-5mm}
\end{figure*}

\subsection{Generation of Accurate Channels}

In Table \ref{tab:channel_dist}, we analyze the ability of our method to capture the underlying distribution of channel matrices compared to baseline methods. To evaluate this, we compare the 2-Wasserstein distance \cite{2wasserstein} and Maximum Mean Discrepancy (MMD) \cite{mmd} between $3000$ samples of generated channels and the ground truth distribution of channels in the testing set.

We observe that the channels generated by our method are closer to the distribution of ground truth channels than those generated by ChannelGAN, which is the best performing baseline, by up to $6 \times$. This shows that our method can generate more realistic channel data as compared to baselines, while ensuring the validity of the generated channels. This is because our method learns to map a known distribution to the distribution over the set of sparse CPD vectors $\mathcal{W}$ as opposed to the full channel matrix $\textbf{H}$, which has no specific sparsity structure and is thus more complex to predict.

\subsection{Cross-Evaluation on Downstream Compression Task}

In Table \ref{tab:cross_eval}, we analyze the efficacy of our method in producing realistic wireless data in the context of a downstream channel compression task, which is crucial to mitigating feedback overheads that arise in practical wireless systems that frequently exchange channel data to optimize communication efficiency  \cite{csinet,csicomp}. We consider two Outdoor scenario datasets for the base stations $10$ and $11$ denoted by $\mathcal{D}^R_{10}$ and $\mathcal{D}^R_{11}$ respectively, to train two generators on each independently. We then use the trained generators to produce synthetic channel datasets, which are denoted by $\mathcal{D}^G_{10}$ and $\mathcal{D}^G_{11}$. Next, we train four autoencoders based on the CSI-Net \cite{csinet} architecture on the datasets $\mathcal{D}^R_{10}$, $\mathcal{D}^R_{11}$, $\mathcal{D}^G_{10}$ and $\mathcal{D}^G_{11}$ respectively, for a compression task using the MSE loss. We then compare the performance of each pairwise combination of autoencoder and dataset. Intuitively, a model trained on $\mathcal{D}^G_{10}$ should generalize well to $\mathcal{D}^R_{10}$, and vice versa. Similar rules apply for models trained on $\mathcal{D}^{R/G}_{11}$. 
In Table \ref{tab:cross_eval}, we observe that an autoencoder trained on data generated by our model follows these rules, indicating that it can capture the distinctions between different datasets in its generated channels. Our method outperforms all baselines, showing that the channels generated by the PPGC pipeline are crucial to downstream compression tasks.

\begin{figure}[t]
    \centering
    \includegraphics[trim={0 0 0 0},clip, width=0.48\columnwidth]{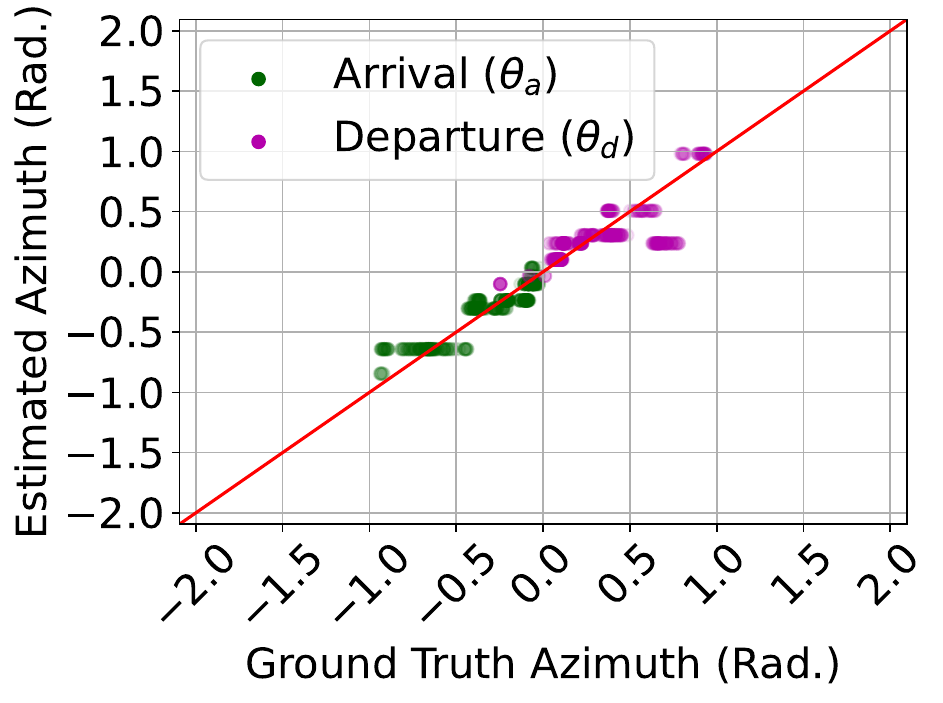}
    \includegraphics[trim={0 0 0 0},clip, width=0.48\columnwidth]{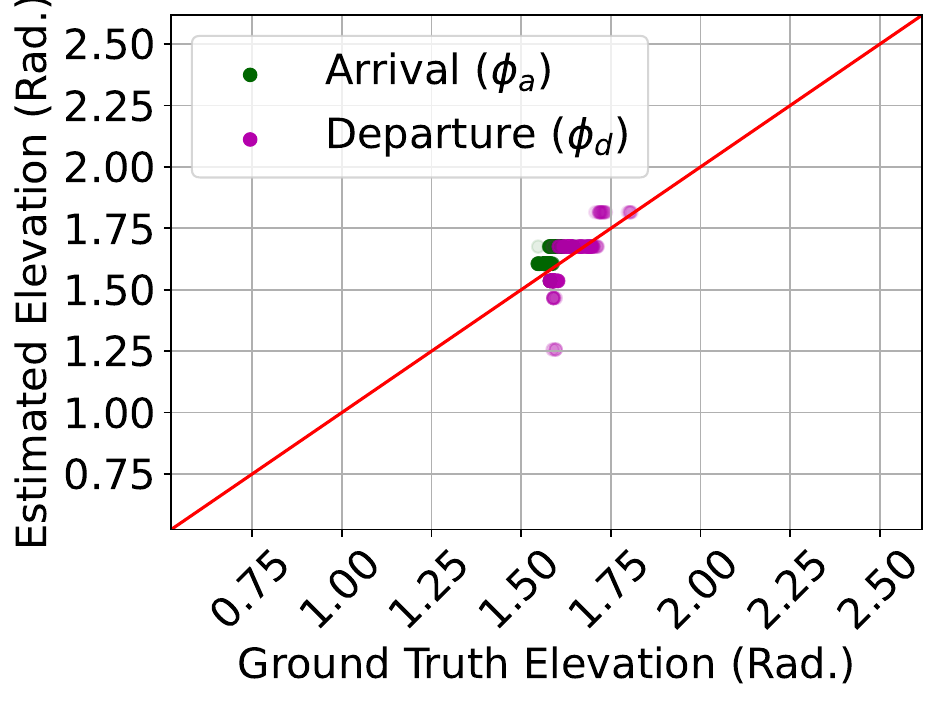}
    \caption{Our method can be extended to estimate channel parameters accurately using the parameter extraction process.}
    \label{fig:exp_parameter_estimation}
    \vspace{-3mm}
\end{figure}

\subsection{Compatibility with Generative Models}

In Fig. \ref{fig:exp_compat}, we illustrate the compatibility of our method with VAE and GAN architectures. As mentioned in Sec. \ref{sec:channel_model}, our method is compatible with  architectures that require no knowledge of the target generator outputs, a condition satisfied by VAE and GAN models, which utilize the MSE loss between the true channel and the output of the complete generator pipeline with no additional constraints. We consider the loss trajectories for the ASU, Boston, and Outdoor datasets, along with samples of the generated channels.

We observe that both VAE and GAN models converge to suitable stationary points as seen in Fig. \ref{fig:exp_compat}, and that the generated channels are very similar to the ground truth. Also, incorporating the response dictionary $\textbf{D}$ guarantees that the generated channels are geometrically valid. Thus, our method is flexible in terms of the choice of the generative model.

\subsection{Accuracy of Parameter Estimation}

In Fig. \ref{fig:exp_parameter_estimation}, we show the compatibility of our framework with autoencoder models for parameter estimation tasks. We use an autoencoder based on the CSI-Net \cite{csinet} architecture, with the output layer producing the set of CPD vectors $\mathcal{W}$. 
We consider the Outdoor dataset with $2$ paths and train for $100$ epochs.

We observe that the autoencoder can accurately predict the underlying channel parameters. This is because the linearized reformulation of the channel model and the decomposition of the parameter tensor allow the autoencoder to converge to optimal parameter values without being affected by the non-convexity of the original channel model $\mathcal{M}$. The discretized nature of the array response dictionary $\textbf{D}$ results in a single estimated angle value for a set of proximal ground truth values. 

\begin{figure*}
   \begin{subfigure}{0.32\textwidth}
        \includegraphics[width=\textwidth]{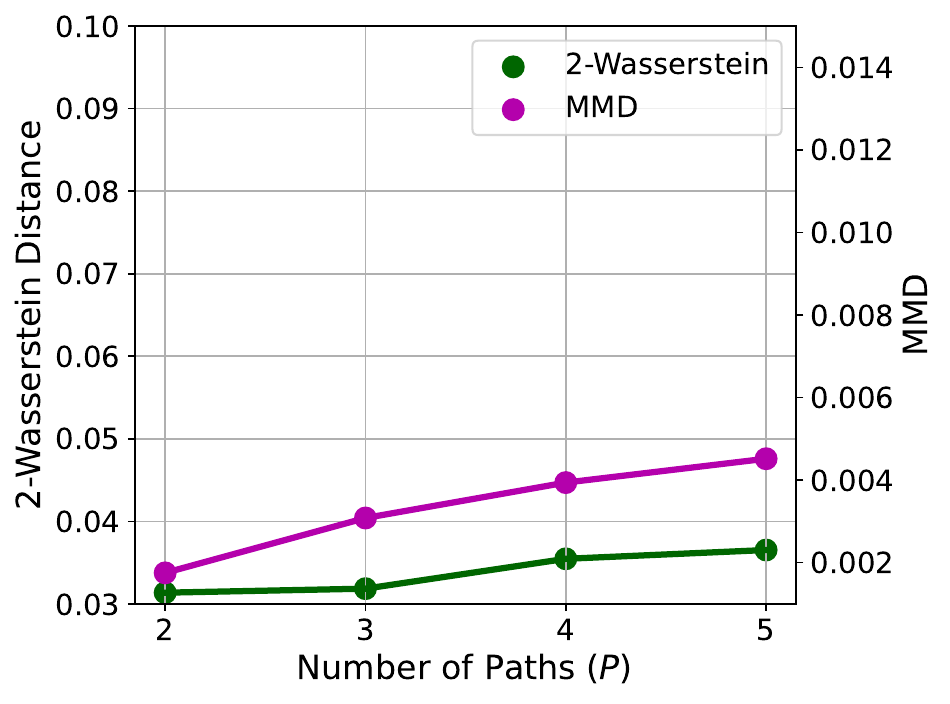}
        \caption{}
        \label{fig:exp_num_paths}
    \end{subfigure}
    \begin{subfigure}{0.32\textwidth}
        \includegraphics[width=\textwidth]{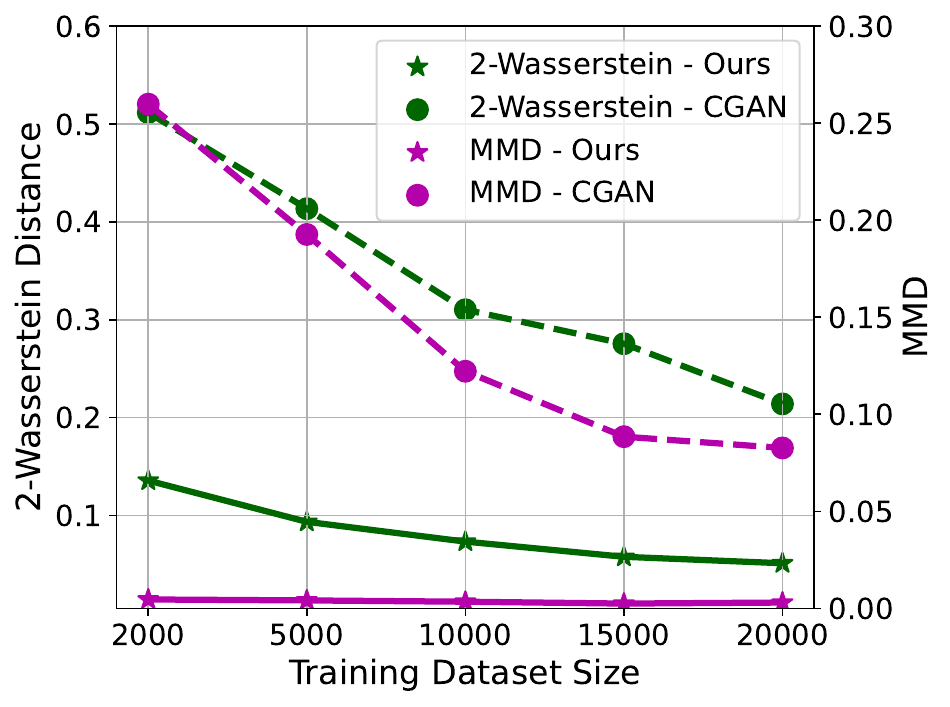}
        \caption{}
        \label{fig:exp_dataset_size}
    \end{subfigure}
    \begin{subfigure}{0.32\textwidth}
        \includegraphics[width=\textwidth]{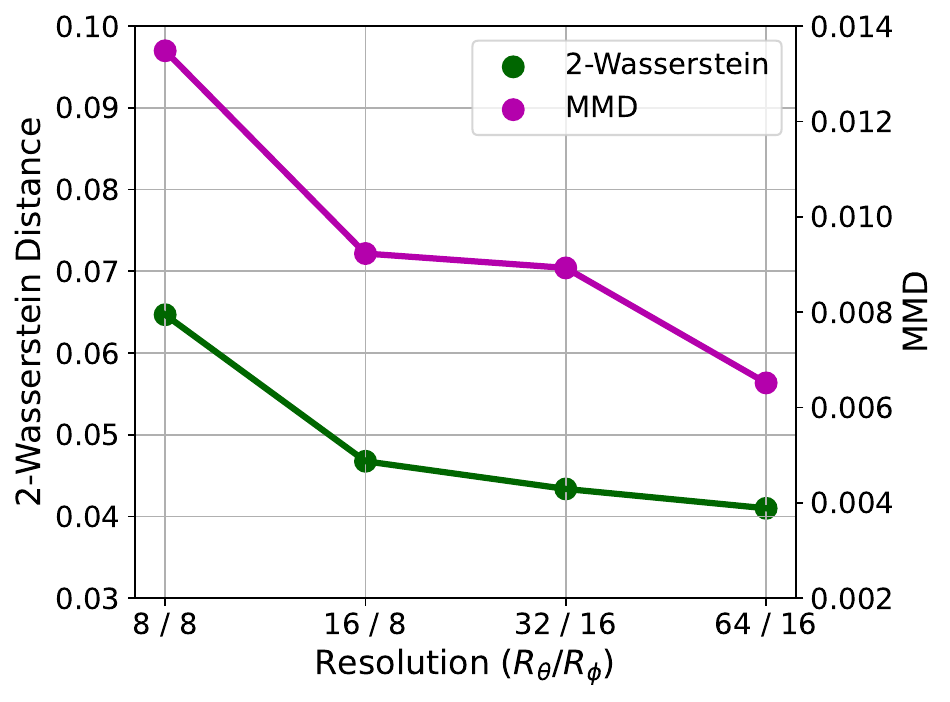}
        \caption{}
        \label{fig:exp_resolution}
    \end{subfigure}
    \caption{Our method generalizes to channel data with an arbitrary number of constituent multipath components while retaining its ability to generate output data that belongs to the distribution of ground truth channels (a), and can generate high fidelity channel data even for smaller sizes of training data compared to the best performing baseline (CGAN). This translates to practical benefits in terms of the resources utilized to gather wireless channel data for training (b). As the resolution $(R_{\theta},R_{\phi})$ of the array response dictionary $\textbf{D}$ is increased, the distribution of channels generated by our method gets closer to the ground truth distribution at the cost of larger computational resources used to store and operate on $\textbf{D}$ (c).}
    \vspace{-5mm}
\end{figure*}

\subsection{Effect of Number of Paths}

In Fig. \ref{fig:exp_num_paths}, we observe the effect of increasing the number of paths $P$ that comprise the input channel $\textbf{H}$ as described in (\ref{eq:channel_sum}). We consider the Indoor dataset with up to $5$ paths, and analyze the ability of our method to adapt to additional paths.

We observe that even as the number of paths $P$ increases, the increase in 2-Wasserstein distance is $\sim 0.002$ on average, and in MMD, it is $\sim 0.0008$ on average for every additional path. This indicates that our method can scale to a larger range of scenarios in terms of the number of constituent paths. This is because the formulation of the generator output assigns a set of independent CPD vectors for each path, allowing the generator to learn the mapping functions for each path independently. It should be noted that the number of paths $P$ affects the size of the output CPD, which is $4RP$ neurons. Thus, the output sizes of the generator are $256,382,512$ and $640$ neurons for $2,3,4$ and $5$ paths, respectively. 

\subsection{Effect of Size of Dataset}

In Fig. \ref{fig:exp_dataset_size}, we observe the effect of varying the amount of training data available. This experiment is motivated by practical considerations in terms of resources expended to gather ground truth wireless data. We consider the Indoor dataset with $5$ paths, which consists of $20,000$ datapoints. We iteratively retrain the generator after removing chunks of the training data, while keeping the validation dataset the same.

We observe that our method retains its ability to generate datapoints similar to the ground truth distribution in terms of 2-Wasserstein distance and MMD even when the size of the dataset is reduced by up to $\sim 50\%$. This illustrates that our method does not require large amounts of training data to produce samples of high-fidelity samples of wireless data. This is because our method learns to map a latent distribution to the distribution over the set of CPD vectors $\mathcal{W}$ rather than the channel matrix $\textbf{H}$, which is an easier mapping to learn on account of the sparsity structure of the CPD vectors $\mathcal{W}.$

\subsection{Effect of Dictionary Resolution}
In Fig. \ref{fig:exp_resolution}, we study the effect of changing the resolution $R_{\theta},R_{\phi}$ of the array response dictionary $\textbf{D}$. We consider the Indoor dataset with $5$ paths, with $(\theta_{\max}-\theta_{\min}) = 2.0$ rad and $(\phi_{\max}-\phi_{\min}) = 1.0$ rad. For resolutions $R_{\theta} = \{8,16\}$, we consider $R_{\phi}=8$ and for resolutions $R_{\theta} = \{32,64\}$, we consider $R_{\phi}=16$. An increase in resolution enables the generative model to capture the underlying parameters of the channel with higher granularity, which results in more accurate generated channel matrices at the cost of an larger dictionary.

We observe that the similarity between generated become and ground truth channels increases with the resolution. This happens because the increased resolution enables the generator to capture the underlying distribution of parameters with greater accuracy, described by the dictionary interval widths $\Delta_{\theta/\phi} = ((\theta/\phi)_{\text{max}}-(\theta/\phi)_{\text{min}})/R_{(\theta/\phi)}$ the azimuth and elevation angles. However, the size of the array response dictionary $\textbf{D}$ increases with resolution as $\mathcal{O}(R_{\theta}^2R_{\phi}^2)$, thus making it more computationally intensive. Thus, a change in resolution constitutes a tradeoff between the quality of the generated channel matrices and the computational resources required to store and operate upon the array response dictionary $\textbf{D}$.

\subsection{Incorporating Additional Parameters}

\begin{figure*}[t]
    \centering
    \begin{subfigure}{0.32\textwidth}
        \includegraphics[trim={0 0 0 0},clip, width=\textwidth]{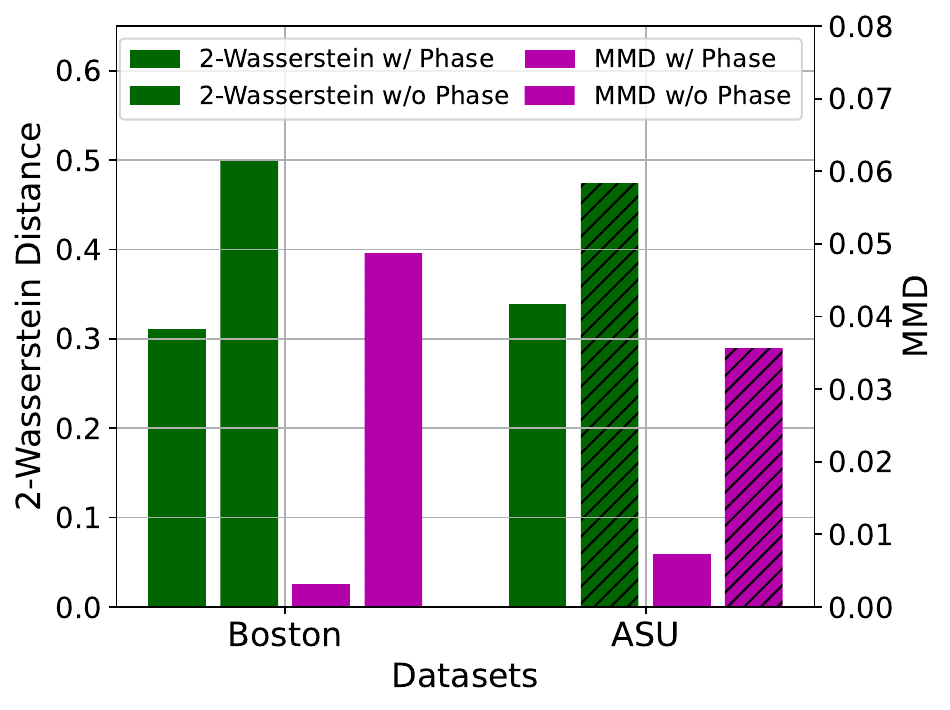}
        \caption{}
        \label{fig:exp_phase}
    \end{subfigure}
    \begin{subfigure}{0.32\textwidth}
        \includegraphics[trim={0 0 0 0},clip, width=\textwidth]{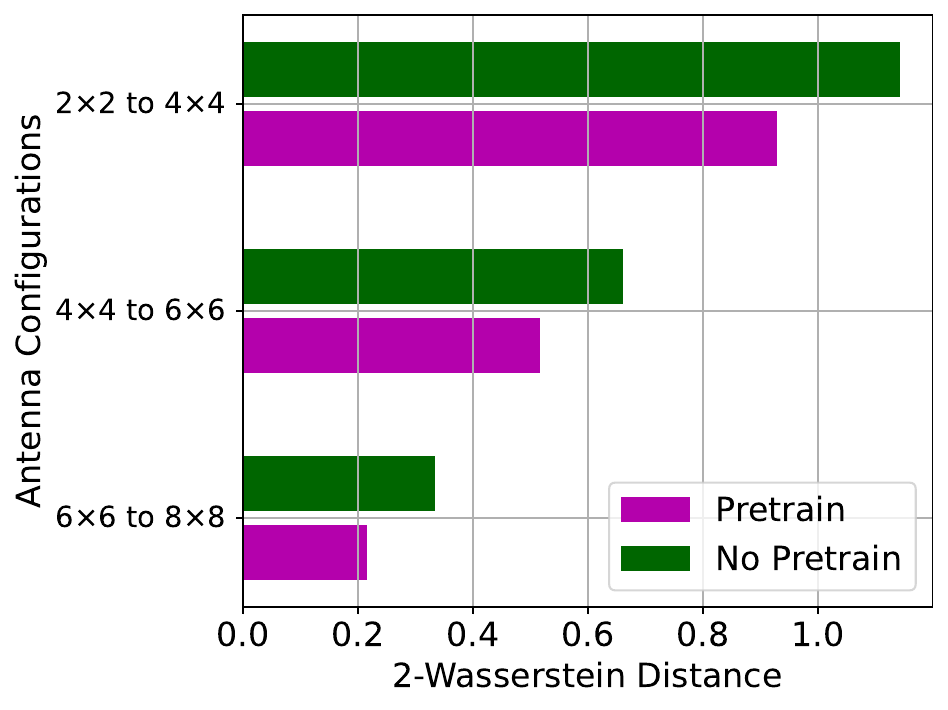}
        \caption{}
        \label{fig:exp_ants_1}
    \end{subfigure}
    \begin{subfigure}{0.32\textwidth}
        \includegraphics[trim={0 0 0 0},clip, width=\textwidth]{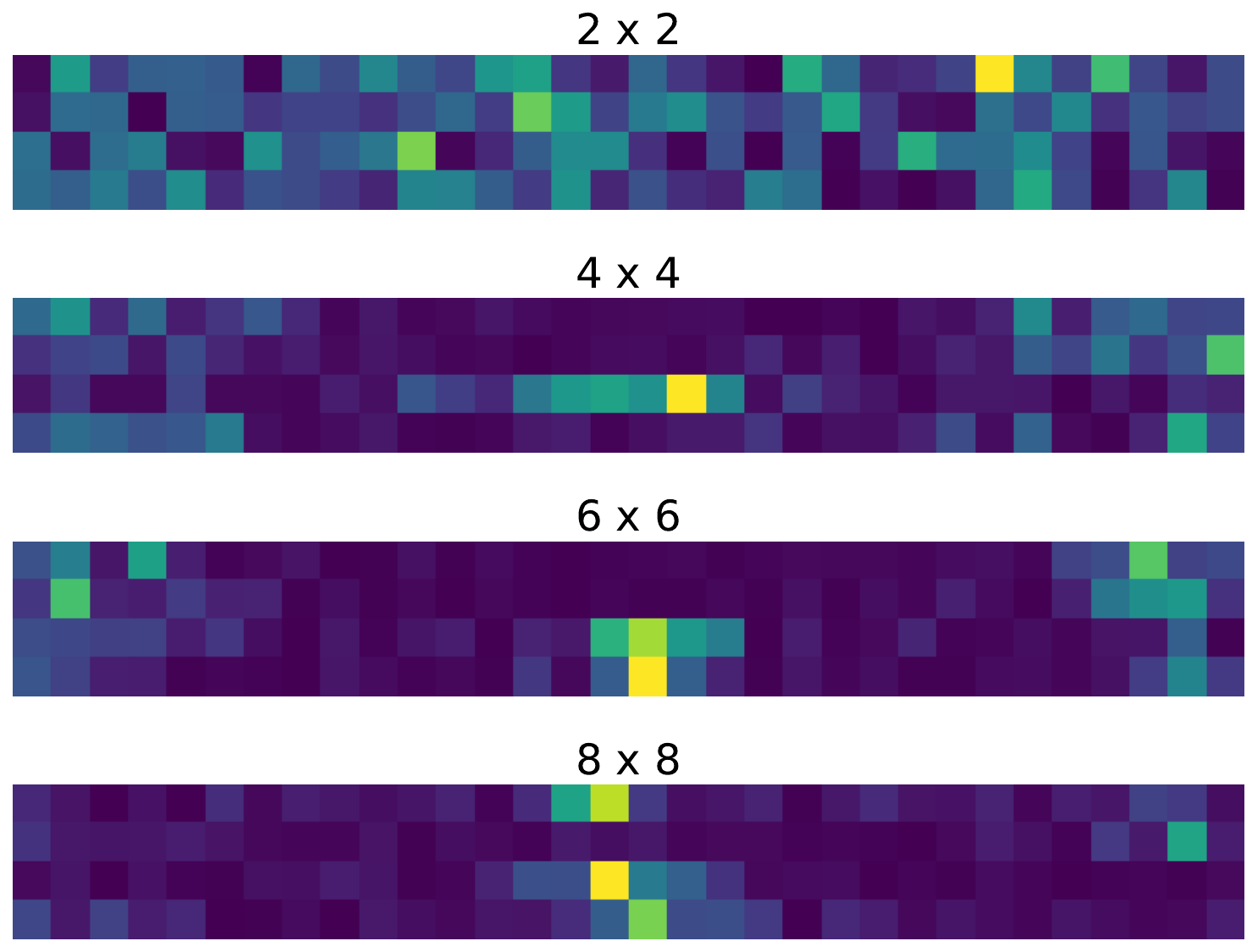}
        \caption{}
        \label{fig:exp_ants_2}
    \end{subfigure}
    \caption{Our method can be extended to incorporate additional model parameters such as channel phase, which results in improved performance on ground truth channel data with a phase component (a), and used as a fine-tuning stage to generalize to novel antenna geometries by leveraging geometry-agnostic multipath information learned by the PPGC (b). Larger antennas allow for finer angular resolution of this multipath information, resulting in more accurate synthetic channels (c).}
    
    \vspace{-3mm}
\end{figure*}

In Fig. \ref{fig:exp_phase}, we demonstrate the ability of our model to incorporate additional channel parameters by including the phase shift caused by the wireless channel as an additional parameter, given by $\{\eta^i\}_{i=1}^P$. We consider the resolution $R_{\eta}=16$ for the phase shift parameter. The output of the generative model is then of size $2R_{\theta}P+2R_{\phi}P+R_{\eta}P$ to account for the additional parameter, resulting in the set of CPD vectors $\mathcal{W}$ consisting of an additional vector for each path corresponding to the phase shift, given by $\{\textbf{w}(\theta_a^i),\textbf{w}(\theta_d^i),\textbf{w}(\phi_a^i),\textbf{w}(\phi_d^i),\textbf{w}(\eta^i)\}$. We evaluate our method on the ASU dataset and the Boston dataset, where the phase component is included in the channel modeling process. We compare the channels synthesized with and without the additional CPD vector for phase and observe the change in 2-Wasserstein distance and MMD.

We observe that our method generates channels that are closer to the ground truth distribution after the addition of the phase shift parameter. This shows that our method can capture the effects of phase shift in addition to the angular parameters through the addition of the corresponding CPD vectors $\{\textbf{w}(\eta^i)\}_{i=1}^P$, which enhances in its ability to synthesize more accurate and realistic distributions of the wireless channel.

\subsection{Transfer Learning across Number of Antennas}

In Fig. \ref{fig:exp_ants_1}, we show that our method can easily be fine-tuned for new antenna-array configurations using a small amount of paired channel data. We pre-train the PPGC model using dataset $\mathcal{D}_{\textsf{Ref}}$ for a reference antenna configuration $N_{r/t}^{\textsf{Ref}}$ and fine-tune on dataset $\mathcal{D}_{\textsf{Paired}}$ of paired reference-target channels, with the reference configuration $N_{r/t}^{\textsf{Ref}}$ and target configuration $N_{r/t}^{\textsf{Trg}}$ similar to \cite{wagle_icassp}. During this phase, we change the precomputed dictionary $\textbf{D}$ to reflect the target configuration with each element of size $N_r^{\textsf{Trg}} \times N_t^{\textsf{Trg}}$.
We compare this method to a PPGC model trained only on  $\mathcal{D}_{\textsf{Paired}}$. We consider three scenarios, namely, $\{N_{r/t}^{\textsf{Ref}}=2/4/6$;$N_{r/t}^{\textsf{Trg}}=4/6/8\}$, $|\mathcal{D}_{\textsf{Paired}}|=50$, and finetune for $100$ minibatch iterations.

We observe that compared to directly training the PPGC on $\mathcal{D}_{\textsf{Paired}}$, our method improves the 2-Wasserstein distance between the synthesized and the ground truth data by up to approximately $30\%$. This indicates that pretraining the PPGC model allows it to learn configuration-agnostic multipath information crucial for adapting to novel antenna geometries. 
We also observe that as the number of antennas in the reference and target configurations become larger, pretraining using our method improves the generation performance to a greater extent. This is because, for larger antenna geometries, critical multipath information can be captured to a higher degree of granularity in the pretraining stage, resulting in the generation of more accurate synthetic channels, as seen in Fig. \ref{fig:exp_ants_2}.

\section{Conclusion}

In this paper, we addressed the problem of validity in channel matrices produced by generative ML models by developing a novel framework that incorporates physics-based channel models in the generative pipeline. We mitigated the convergence issues faced in the straightforward incorporation of channel models by relaxing the non-convexity of the model through a discretized array response dictionary which simplifies it to a linear combination problem. We also improved the scaling characteristics of our method by generating the tensor decomposition of the overall parameter tensor, enabling it to scale to larger resolutions and number of parameters. Extensive experimental validation comparing generated channel distributions as well as performance on downstream tasks shows that the proposed method outperforms the different state-of-the-art methods for generative channel modeling. Future directions include temporal modeling of channels, increasing the number of parameters to better capture the behavior of the channels.

\bibliographystyle{IEEEtran}
\bibliography{refs}

\end{document}